\documentclass[fleqn,usenatbib]{mnras}

\usepackage{newtxtext,newtxmath}
\usepackage[T1]{fontenc}

\DeclareRobustCommand{\VAN}[3]{#2}
\let\VANthebibliography\thebibliography
\def\thebibliography{\DeclareRobustCommand{\VAN}[3]{##3}\VANthebibliography}

\usepackage{graphicx}	% Including figure files
\usepackage{amsmath}	% Advanced maths commands
\usepackage{bm}
\usepackage{float}
\title[SSA of low frequency radiometric data]{Singular spectrum analysis of time series data from low frequency radiometers, with an application to SITARA data}

\author[J. N. Thekkeppattu et al.]{
Jishnu N. Thekkeppattu,$^{1,2}$\thanks{E-mail: j.thekkeppattu@curtin.edu.au (JNT)}
Cathryn M. Trott,$^{1,3}$
Benjamin McKinley$^{1,3}$
\\
$^{1}$International Centre for Radio Astronomy Research, Curtin University, Bentley, WA 6102, Australia\\
$^{2}$Raman Research Institute, C V Raman Avenue, Sadashivanagar, Bengaluru 560080, India\\
$^{3}$ARC Centre of Excellence for All Sky Astrophysics in 3 Dimensions (ASTRO 3D), Bentley, WA 6102, Australia
}

\date{Accepted XXX. Received YYY; in original form ZZZ}

\pubyear{2022}

\begin{document}
\label{firstpage}
\pagerange{\pageref{firstpage}--\pageref{lastpage}}
\maketitle

\begin{abstract}
Understanding the temporal characteristics of data from low frequency radio telescopes is of importance in devising suitable calibration strategies. Application of time series analysis techniques to data from radio telescopes can reveal a wealth of information that can aid in calibration. In this paper, we investigate singular spectrum analysis (SSA) as an analysis tool for radio data. We show the intimate connection between SSA and Fourier techniques. We develop the relevant mathematics starting with an idealised periodic dataset and proceeding to include various non-ideal behaviours. We propose a novel technique to obtain long-term gain changes in data, leveraging the periodicity arising from sky drift through the antenna beams. We also simulate several plausible scenarios and apply the techniques to a 30-day time series data collected during June 2021 from SITARA - a short-spacing two element interferometer for global 21--cm detection. Applying the techniques to real data, we find that the first reconstructed component - the trend - has a strong anti-correlation with the local temperature suggesting temperature fluctuations as the most likely origin for the observed variations in the data. We also study the limitations of the calibration in the presence of diurnal gain variations and find that such variations are the likely impediment to calibrating SITARA data with SSA.
\end{abstract}


%%%%%%%%%%%%%%%%%%%%%%%%%%%%%%%%%%%%%%%%%%%%%%%%%%

%%%%%%%%%%%%%%%%% BODY OF PAPER %%%%%%%%%%%%%%%%%%

\section{Introduction}

There is a renewed interest in low-frequency (< 300~MHz) radio astronomy, due to the multitude of science cases that benefit from low-frequency observations. Several low-frequency radio telescopes such as LOFAR  \citep{2013A&A...556A...2V}, MWA \citep{2013PASA...30....7T}, HERA \citep{DeBoer_2017} and  LWA\citep{2015AAS...22532801H} have been constructed and are currently   observing, with the low frequency Square Kilometer Array, SKA-low \citep{2009IEEEP..97.1482D} in construction phase. Some of the key science goals for these metre wavelengths radio telescopes are cosmic dawn and epoch of reionisation (CD/EoR)\citep{trott_2017}, solar and heliospheric science \citep{NINDOS20191404} and cosmic magnetism \citep{GAENSLER20041003}. Modern low frequency radio telescopes differ from their higher frequency counterparts in that they consist of aperture arrays constructed from large numbers of antennas, that are often beamformed in the analog or digital domain and correlated against each other to observe the radio sky. The calibration requirements, calibration models and the complexity are different to higher frequency (cm wavelengths and above) radio telescopes.

For dish-based interferometers with a small number of antennas, single dish telescopes and specialised low frequency radiometers, calibration techniques such as Dicke switching or noise injection can be employed. Specifically, global 21--cm experiments constitute a group of low-frequency radiometers requiring precise and accurate calibration of the systems to limit the systematics to less than one part in a million. Most of the global 21--cm experiments use single antennas as sensors and employ Dicke-switching ambient temperature thermal loads and noise diodes (see \cite{https://doi.org/10.1029/2011RS004962,T.2021}) or noise injection \citep{2018ExA....45..269S} for bandpass calibration. The same calibrators are used to compensate for receiver gain drifts over extended periods of time. However, in order to maintain a stable excess noise ratio (ENR), the calibrators have to be maintained in temperature controlled environments, and the noise diodes themselves require periodic re-calibration with laboratory standards to mitigate drift and ageing, especially if they are to be used in applications demanding high accuracy such as cosmology. Besides, with future telescopes such as the SKA-low potentially employing thousands of low cost active antennas, calibration using dedicated noise diodes at each antenna becomes impossible. The aperture array nature of these instruments, i.e. large numbers of stationary antennas of simple construction, often with integrated low-cost low noise amplifiers (LNAs), necessitates calibration based on sky models. Therefore, development of novel mathematical tools to explore the long-term stability of these low-frequency telescopes and radiometers, and to determine the limits of calibration and data integration, become essential.

In this paper, we explore the potential of one such tool in analysing time series data from low frequency radiometers, and an application of it to real valued time series data. The instrument under study is SITARA, a broadband two element interferometer targeting global 21--cm detection employing short spacing interferometry. SITARA consists of two MWA style active antennas kept 1~m apart over a large ground-plane (35~m) and a correlator that records auto-correlations and cross-correlations with spectral resolution of about 61~kHz, across 0--250~MHz. The usable band is limited to 70--200~MHz, with reduced sensitivity in the 50--70~MHz band. SITARA is deployed within the radio quiet zone of the Murchison Radioastronomy Observatory (MRO) in Western Australia, which is also the site for future SKA-low. Data are collected round the clock and timestamped data are written out in \texttt{miriad} format \citep{1995ASPC...77..433S}. SITARA has been conceived as the first prototype to evolve tools and techniques for short-spacing interferometry; further details about SITARA can be found in \cite{2022PASA...39...18T}.
Single frequency data from SITARA auto-correlations can essentially be treated as radiometric data, with SITARA behaving as an uncalibrated total power radiometer. 

\section{Motivation}
\begin{figure}
	\includegraphics[width=\columnwidth]{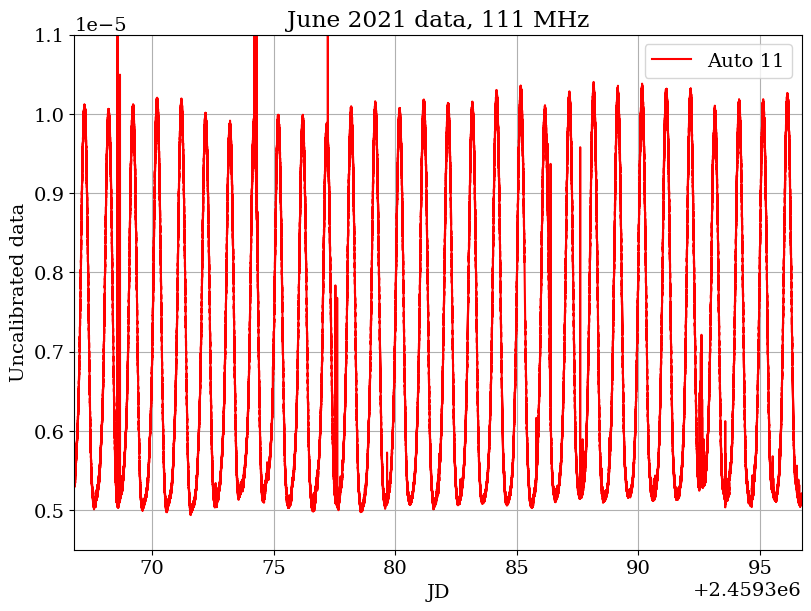}
    \caption{SITARA raw data time series as a function of Julian Date (JD) for a single frequency channel of bandwidth $61$~kHz, at a frequency of 111~MHz. The data have been extracted from a concatenated SITARA dataset for the month of June 2021. Solar bursts contribute to most of the RFI seen in this time series.}
    \label{fig:raw_data_111MHz}
\end{figure}

As an example to motivate this study, we consider real valued auto-correlations from SITARA data. A plot of single frequency channel time series data from SITARA auto-correlations, at a frequency of 111~MHz for the month of June 2021, is shown in Fig.\ref{fig:raw_data_111MHz}. The frequency of 111~MHz has relatively low radio frequency interference (RFI) and is also a frequency where mutual coupling between antennas does not cause large beam shifts (see \cite{2022PASA...39...18T} for details). The data have a periodic nature arising from the drifting of various regions of the radio sky through the antenna beams as the Earth rotates. If the radio telescope was perfectly calibrated, and there was no RFI, Fig.\ref{fig:raw_data_111MHz} would have shown a perfectly repeating pattern. 

As can be seen, this is not the case. The data have some multiplicative gain variations as well as some additive RFI. Therefore, our aim is to decompose this time series into some components that can help us understand it better. A plausible decomposition is the Fourier transform, i.e. into sines and cosines. However, radiometric data can have aperiodic patterns such as  a drift which gets distributed to various modes in a Fourier analysis, thus making an interpretation difficult. Therefore, it is desirable to adopt an analysis which selects a suitable basis from the data; in other words a data driven technique. 

Singular spectrum analysis (SSA) is a set of data driven tools that can decompose a time series into elementary patterns such as trend and oscillatory components. The raw data from a low frequency radio telescope with a fixed pointing are expected to have oscillatory components with a period corresponding to a sidereal day, while owing to environmental changes trend-like patterns (drift) are also expected. This makes SSA an ideal tool for analysis of such data. 

Singular spectrum analysis (SSA) techniques appear in the analysis of dynamical systems \citep{BROOMHEAD1986217, VAUTARD1989395} and SSA has been a popular tool for time series analysis in a variety of fields such as meteorology and climate science \citep{Ghil20021} and geophysics \citep{10.1093/gji/ggw473}. However the application of SSA in radio astronomy has been limited \citep{2016Ap.....59..199D, 8575420}. Indeed, to the best of our knowledge, SSA has not been applied for time series analysis of radiometric time series data. This paper aims to detail the necessary mathematical tools for SSA of time series data from a low frequency radiometer and demonstrate them with SITARA data.

\subsection{Notation and mathematical preliminaries}
In this section, we describe the notation employed and define certain matrices that are useful for the subsequent analysis. Many of these definitions can be found in \cite{Davis_circulant} as well as \cite{10.1115/1.4027722}. We use bold capital Roman letters such as $\bm{X}$ to denote matrices, and small Roman letters such as $n$ are used for indexing. We use zero-based numbering such that the indices start at $0$. Small Roman letters with an arrow such as $\vec{u}$ denote vectors. Integers are denoted by capital Roman letters such as $L$. 
We now proceed to define some basic matrices and associated linear algebra.
\subsubsection{Circulant and anti-circulant matrices}
An $N \times N$ square matrix $\bm{C}$ is circulant if each row of the matrix is a \textit{right} shifted version of the previous row as shown in Eq.\ref{Eq:C_circ}.
\begin{equation}
\label{Eq:C_circ}
\bm{C} = 
\begin{bmatrix}
c[0] & c[1] &...... & c[N-1] \\
c[N-1] & c[0] &...... & c[N-2] \\
c[N-2] & c[N-1] &...... & c[N-3] \\
..  & ..  &...... & .. \\
c[1] & c[2] &...... & c[0]\\
\end{bmatrix}\\
\end{equation}

An $N \times N$ square matrix $\bm{C}$ is anti-circulant if each row of the matrix is a \textit{left} shifted version of the previous row as shown in Eq.\ref{Eq:C_acirc}.
\begin{equation}
\label{Eq:C_acirc}
\bm{C_a} = 
\begin{bmatrix}
c[0] & c[1] &...... & c[N-1] \\
c[1] & c[2] &...... & c[0] \\
c[2] & c[3] &...... & c[1] \\
..  & ..  &...... & .. \\
c[N-1] & c[0] &...... & c[N-2]\\
\end{bmatrix}\\
\end{equation}
Both the circulant and anti-circulant matrices can be obtained from a sequence $c[n]; n=0,1,....N-1$ and are completely specified by that sequence. While useful in a wide variety of analyses, circulant matrices are not of much utility in this paper. We will be dealing with anti-circulant matrices instead.
A matrix can also be interpreted to have been broken into blocks or submatrices which are themselves matrices.
\subsubsection{Matrix products}
The notations employed in this paper for the various matrix products are listed below. 
\begin{enumerate}
    \item Regular matrix-matrix multiplication is denoted with no specific operator.
    % \item $\otimes_{K}$ denotes the Kronecker product between two matrices. 
    \item $\otimes$ denotes the outer product of two vectors that results in a matrix.
    \item $\odot$ represents an element-by-element multiplication of two matrices, known as the Hadamard product.
\end{enumerate}

\section{Basics of singular spectrum analysis}

Though the theory of SSA is covered in detail in references such as \cite{golyandina2001analysis}, we provide a basic description of the steps involved for completeness. In this paper we follow the SSA approach known as the Broomhead-King (BK) version, with the alternate being the Vautard-Ghil (VG) version. The VG version is only suitable for the analysis of a stationary time series and therefore is not discussed here. Following \cite{GOLYANDINA2014934}, the four major steps in BK SSA are given below. 

\begin{enumerate}
    \item Convert the 1-D time series into a 2-D matrix called a \textit{trajectory matrix}. This step is called \textit{embedding} in time series analysis.
    \item Decompose the trajectory matrix with singular value decomposition (SVD). The result consists of a set of left and right singular vectors and associated singular values. 
    \item Reconstruct the constituent components of the trajectory matrix with selected singular vectors and singular values.
    \item Reconstruct the time series with these components by performing diagonal averaging.
\end{enumerate}

In the embedding step, the trajectory matrix is constructed with columns consisting of elements from sliding a window of length $L$ across the original time series.  The window length is chosen in such a manner that each column of the resulting matrix consists of one cycle of the data. As an example, for drift sky data from a ground based radiometer, as given in Sec.\ref{subsec:periodic_sims} this corresponds to one sidereal day . For each sliding, the elements inside the window are made into one column of the trajectory matrix, yielding a matrix with $L$ rows and $K = N-L+1$ columns. Consider a time series $x[n], n=0,1,...N-1$. The embedding step converts this time series of length $N$ into an $L\times K$ matrix that has the elements of the time series as given in Eq.\ref{Eq:traj_matrix}. 
\begin{equation}
\label{Eq:traj_matrix}
\bm{X} = 
\begin{bmatrix}
x[0] & x[1] & x[2] & .. & x[K-1] \\
x[1] & x[2] & x[3] & .. & x[K] \\
x[2] & x[3] & x[4] & .. & x[K+1] \\
..  & .. & .. & .. & .. \\
x[L-1] & x[L] & x[L+1] & .. & x[N-1]\\
\end{bmatrix}\\
\end{equation}
From Eq.\ref{Eq:traj_matrix}, it can be seen that the anti-diagonals of a trajectory matrix contain similar terms, and therefore it is like a Hankel matrix, although it is not square in general. It may be noted that some implementations pad the original time series with zeros to obtain $K=N$, though we do not employ this. 

In the second step, the trajectory matrix is decomposed via SVD to yield left and right singular vectors as well as the corresponding singular values. This can be written as 
\begin{equation}
    \bm{X} = \bm{U}\bm{\Sigma} \bm{V}^T.
\end{equation}
The SVD operation decomposes the $L\times K$ matrix $\bm{X}$ into three matrices $\bm{U},\bm{\Sigma}$ and $\bm{V}$; where $\bm{U}$ is an $L\times L$ unitary matrix, $\bm{V}$ is a $K \times K$ unitary matrix and $\bm{\Sigma}$ is an $L \times K$ diagonal matrix consisting of the singular values. Since we consider only real valued matrices $\bm{X}$ in this paper, the matrices $\bm{U}$ and $\bm{V}$ are real orthogonal matrices. Each singular value $\sigma_i$ and the corresponding singular vectors $\vec{u_i}$ and $\vec{v_i}$ form an eigentriple $(\sigma_i, \vec{u_i}, \vec{v_i})$. The decomposition step can also be written as 
\begin{equation}
\label{Eq:X_expansion}
    \bm{X} = \sum_i \bm{X_i}; \;
    \bm{X_i} = \sigma_i (\vec{u_i}\otimes \vec{v_i}). 
\end{equation}
where $\otimes$ is the vector outer product of two vectors yielding a matrix. 

These eigentriples can be grouped and used to reconstruct the various components of the time series. An inspection of the singular values can give an insight into the complexity of the data and the number of eigentriples to consider, and this is also related to the rank of the trajectory matrix.  It may be noted that calculation of SVD of a data matrix and subsequent reconstruction is identical to principal component analysis or PCA \citep{1986pca..book.....J}. Seen in this light, SSA can be treated as PCA of the trajectory matrix formed out of a time series. However, PCA does not involve embedding or trajectory matrices, and therefore the interpretation of the PCA results is different to that of SSA. 

The next step is to reconstruct the trajectory matrix with the selected eigentriples. 
Once this is accomplished, a reconstructed time series is obtained by performing an anti-diagonal averaging over the reconstructed matrix - this step is called ``diagonal averaging'' \citep{GOLYANDINA2014934} \footnote{The procedure is technically anti-diagonal averaging, however we use the term ``diagonal averaging'' to be consistent with the SSA literature.}. The diagonal averaging step applied to a trajectory matrix $\bm{Y}$ yields a series $y_s$ as shown in Eq.\ref{Eq:diag_av}.
\begin{equation}
    \label{Eq:diag_av}
    y_s[n] = \frac{\sum_{(l,k) \in A_s} \bm{Y}[l][k]}{|\mathbb{A}_n|}
\end{equation}
where $\mathbb{A}_n = \{(l,k); l+k = n, 0 \leq l \leq L-1, 0 \leq k \leq K-1\}$ and $|\mathbb{A}_n|$ is the number of elements in the set $\mathbb{A}_n$.
This step can be treated as the reverse of the embedding step, conversion of an anti-diagonal, Hankel-like matrix back into a time series. To gain an appreciation of this step, it is useful to take a closer look at the structure of a trajectory matrix as given in Eq.\ref{Eq:traj_matrix}. Recalling that this matrix is formed by sliding a window and converting the data into columns of a matrix, it can be seen that the anti-diagonals consist of the same value from the time series. Therefore to convert a trajectory matrix back into a time series, all that one has to do is a computation of the mean of the redundant information in each anti-diagonal and form a time series out of them. This operation is what Eq.\ref{Eq:diag_av} does. It may also be noted that one can even select the first row and the last column of a trajectory matrix to reconstruct a time series, however it is preferable to obtain a mean of the redundant information in the anti-diagonal elements. When applied to each $\bm{X_i}$ in Eq.\ref{Eq:X_expansion}, diagonal averaging results in \textit{reconstructed series}. However, one can also choose to form $\bm{X_i}$ from groups of eigentriples, and then obtain reconstructed series. For example, one may pair the eigentriples, then each eigentriple used to form a matrix (by forming outer products and scaling with the corresponding singular value) and a pair of matrices added together to form one $\bm{X_i}$. Indeed, this grouping approach is what we employ in this paper. Therefore, the result is the decomposition of the original time series into a sum of reconstructed series as shown in Eq.\ref{eq: recon_series_eq}.
\begin{equation}
\label{eq: recon_series_eq}
    x[n] = \sum_{i=0}^{h} \widetilde{x_i}[n], n=0,1,...N-1 
\end{equation}
where  $\widetilde{x_i}$ is the reconstructed series obtained from $\bm{X_i}$  and $h$ depends on the grouping of eigentriples. For elementary grouping, $h=L-1$ as there can only be a maximum of $L$ singular values for a matrix of dimensions $L \times K$. Depending on the selection of eigentriples, the reconstructed time series reveals the corresponding aspect of the data. For example, the first eigentriple contains information on the trend, while the subsequent ones are associated with oscillatory patterns. Higher eigentriples are typically associated with noise. 

\section{SSA of periodic time series}
\label{sec:SSA_periodic}
In this section, we obtain the mathematical form of SSA when applied to a time series containing periodic data. We first consider the case where we have an ideal periodic time series. We find that in this case, the singular vectors obtained are sinusoidal in nature. The case where a periodic series is corrupted by multiplicative element is then considered. We also perform simulations to validate the algebra.
\subsection{SSA applied to an ideal periodic time series}
Consider a time series $x[n], n=0,1,....N-1$ of length $N$ that is strictly periodic with a period given by $L$. 
\begin{equation}
    x[k + pL] = x[k]
\end{equation}
We also assume that the length $N$ of the time series obeys the relation $N = zL-1$, where $z$ is an integer. Forming a trajectory matrix out of this sequence gives the following matrix.
\begin{equation}
\label{Eq:traj_matrix[per}
\bm{X} = 
\begin{bmatrix}
x[0] & x[1] &...... & x[L-1] \\
x[1] & x[2] &...... & x[0] \\
..  & ..  &...... & .. \\
x[L-1] & x[0] &...... & x[L-2]\\
\end{bmatrix}_{L \times K}\\
\end{equation}
As $K$ is an integer multiple of $L$, the trajectory matrix can be interpreted as a block matrix, and partitioned into $M = \frac{K}{L}$ square submatrices as given in Eq.\ref{Eq:sq_smatrix}, where $\bm{X_{ac}}$ is an anti-circulant matrix of dimensions $L\times L$.
\begin{align}
\label{Eq:sq_smatrix}
\bm{X} &= \begin{bmatrix}
\bm{X_{ac}}~~\bm{X_{ac}} ..
\end{bmatrix}_{L\times M} \\
\bm{X_{ac}} &= 
\begin{bmatrix}
x[0] & x[1] &...... & x[L-1] \\
x[1] & x[2] &...... & x[0] \\
x[2] & x[3] &...... & x[1] \\
..  & ..  &...... & .. \\
x[L-1] & x[0] &...... & x[L-2]\\
\end{bmatrix}
\end{align}
Before proceeding further, some observations can be made.
\begin{enumerate}
    \item We find that $rank(\bm{X}) \le L$, as there can only be a maximum of $L$ linearly independent columns in $\bm{X}$. Consequently there can be a maximum of $L$ non-zero singular values for $\bm{X}$.
    \item The anti-circulant matrix form given in Eq.\ref{Eq:sq_smatrix} is real symmetric. Therefore, it has an eigendecomposition given as $\bm{X_{ac}} = \bm{Q \Lambda Q^T}$,  where $\bm{Q}$ is an $L\times L$ orthonormal matrix of eigenvectors and $\bm{\Lambda}$ is an $L\times L$ matrix of the corresponding eigenvalues. 
\end{enumerate} 
This enables us to write the trajectory matrix as a  block matrix $\bm{X}=\begin{bmatrix}
\bm{Q \Lambda Q^T}~~\bm{Q \Lambda Q^T} ..
\end{bmatrix} $. Without going through the pedagogical details, we state that $\bm{X}$ can be decomposed into three matrices as given below.
 
\begin{align}
\label{Eq:SVD_eig_expansion}
    \bm{X} &= \bm{Q}_{L\times L}
        \begin{bmatrix} \sqrt{M}\bm{\Lambda} & \bm{0}  & .. \end{bmatrix}_{L\times K}
        \frac{1}{\sqrt{M}}\begin{bmatrix} \bm{Q^T} & \bm{Q^T} & ..\\
        \bm{Q^T} & \bm{Q^T} & ..\\
        .. & .. & .. \\
        \end{bmatrix}_{K\times K} \\
         &= \bm{U}_{L\times L}\bm{\Sigma}_{L\times K} \bm{V^T}_{K\times K}
\end{align}

where, in the last step we recall the SVD of $\bm{X} $ to facilitate a direct comparison. 

 We now attempt to find the nature of this decomposition - specifically what one can expect from the singular vectors if the time series is periodic in nature. From Eq.\ref{Eq:SVD_eig_expansion}, it can be noted that the singular values are a scaled version of the eigenvalues of anti-circulant matrix $\bm{X_{ac}}$. Inspecting the block matrix representation in Eq.\ref{Eq:SVD_eig_expansion}, it can also be seen that the matrix that forms to the left and right singular vectors is the matrix of eigenvectors $\bm{Q}$. Therefore, all that we need to know is the nature of $\bm{Q}$. A calculation of this is provided in Appendix.\ref{Ap:acirc_eigen}, where it is shown that $\bm{Q}$ consists of sinusoidal eigenvectors if the data are periodic. Therefore, the left singular vectors - which are exactly the eigenvectors contained in the matrix $\bm{Q}$ - are all sinusoidal. Besides, the vectors in $\bm{Q}$ are all periodic in $L$ and thus when arranged as blocks in the right singular vector matrix, they form continuous sinusoids. Thus, the equivalence of decompositions in Eq.\ref{Eq:SVD_eig_expansion} shows that \textit{when SSA is applied to a strictly periodic sequence, the singular vectors obtained are sinusoidal}. 

In Appendix.\ref{Ap:acirc_eigen}, it is also shown that the eigendecomposition of anti-circulant matrices such as  $\bm{X_{ac}}$ has a deep relation to discrete Fourier transforms (DFT). Indeed, when applied to an ideal periodic series, SSA results are directly comparable to DFT of one cycle of periodic data; this is also demonstrated with simulations in Sec.\ref{subsec:periodic_sims}. However, if the data have aperiodic structures, DFT of the entire span of data is not a suitable choice as aperiodic structures are not well represented in a sine/cosine Fourier basis. In such a scenario, SSA is better suited to reveal aperiodic structures in the data. Therefore, regardless of the nature of data, SSA can be applied to them and if the data are periodic, SSA reduces to a form of Fourier analysis. 

Even though the decomposition given may be treated as an SVD of the trajectory matrix, some caution has to be exercised.  Since the singular vector matrices have to be unitary, the right singular matrix gets divided by a scaling factor of $\sqrt{\frac{K}{L}}$  while the singular values get multiplied the same factor. Also, the left singular matrix in SVD is unitary, while the eigenvector form given in Appendix.\ref{Ap:acirc_eigen} has unit amplitude. The corresponding scaling applies to the singular value, however it is inconsequential when the eigendecomposition and SVD are computed with numerical packages. Moreover, eigenvalues can be positive and negative, and as given in Appendix.\ref{Ap:acirc_eigen}, they occur in positive and negative pairs in this context. However, singular values are always non-negative and therefore the signs of the eigenvalues get  applied to  the singular vectors when equated with the SVD given in Eq.\ref{Eq:SVD_eig_expansion}. Also, we assume that both the singular values and eigenvalues (and corresponding vectors) have been ordered in the same fashion - typically in descending magnitude. 

\subsubsection{Simulations}
\label{subsec:periodic_sims}
To validate the above calculations, we analyse a simulated periodic time series. Since our aim is to gain a better understanding of SSA as applied to radiometric time series data, the simulations are beam multiplied sky temperature as would be seen by a radiometer. The simulation methodology used to obtain the time series is described in \cite{2022PASA...39...18T}, which we have extended to yield a time series spanning 30 days. For this, we make an important assumption that the radio sky is static and therefore, the true sky temperature is exactly the same across all days for each local sidereal time. The simulations are for auto-correlations at a frequency of 111~MHz with a time cadence of 15 minutes of sidereal time. Since the sky has a periodicity of one sidereal day, it is important that the cadence is chosen in sidereal time units. 

\begin{figure}
\centering
	\includegraphics[width=\columnwidth]{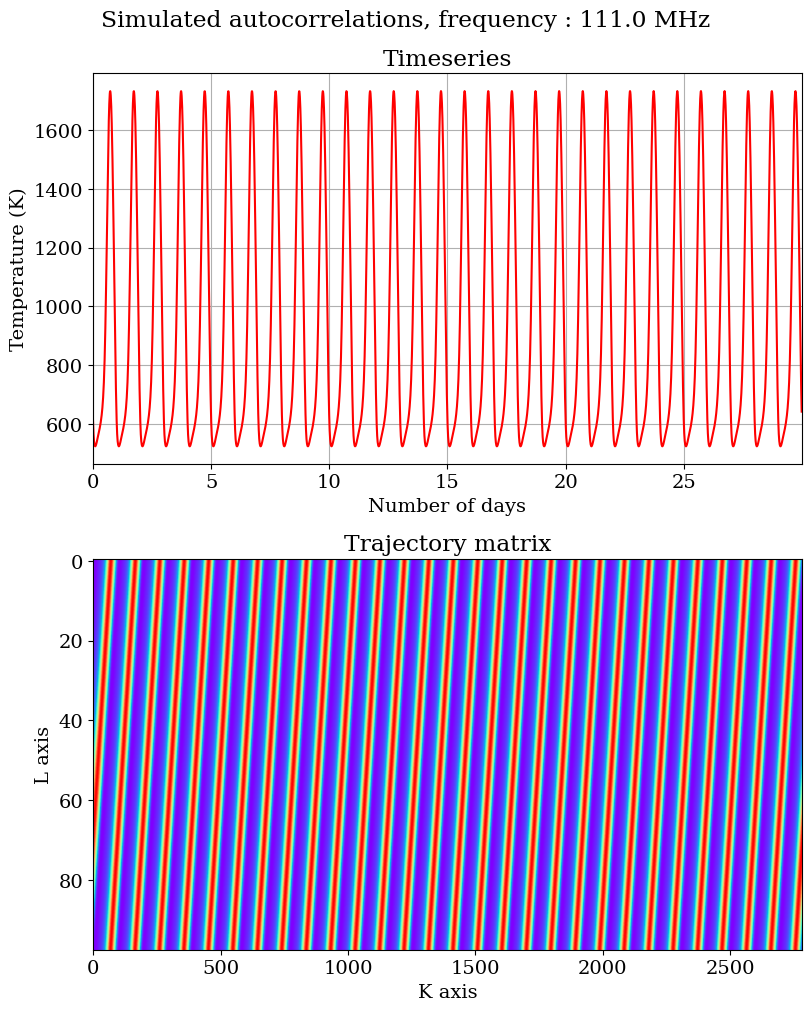}
    \caption{Simulated time series at a frequency of 111~MHz and the corresponding trajectory matrix formed by choosing an embedding dimension $L=96$.}
    \label{fig:simu_111MHz}
\end{figure}

The simulation yields $N=2879$ data points which are then converted into a trajectory matrix.
Key to embedding data is selection of an appropriate window length. Typically, a window length is chosen such that it is divisible by the fundamental of the known periodicity. As we have prior information that data have a periodicity corresponding to a sidereal day, a choice of the window length is the number of samples that correspond to one sidereal day. Therefore a suitable embedding dimension is $L=96$; as it corresponds to one sidereal day. This gives a matrix of dimensions $96 \times 2784$ which can be partitioned into $M = \frac{K}{L} = 29$ block matrices. The simulated time series and a representation of the trajectory matrix are shown in Fig.\ref{fig:simu_111MHz}. 

\begin{figure}
\centering
	\includegraphics[width=\columnwidth]{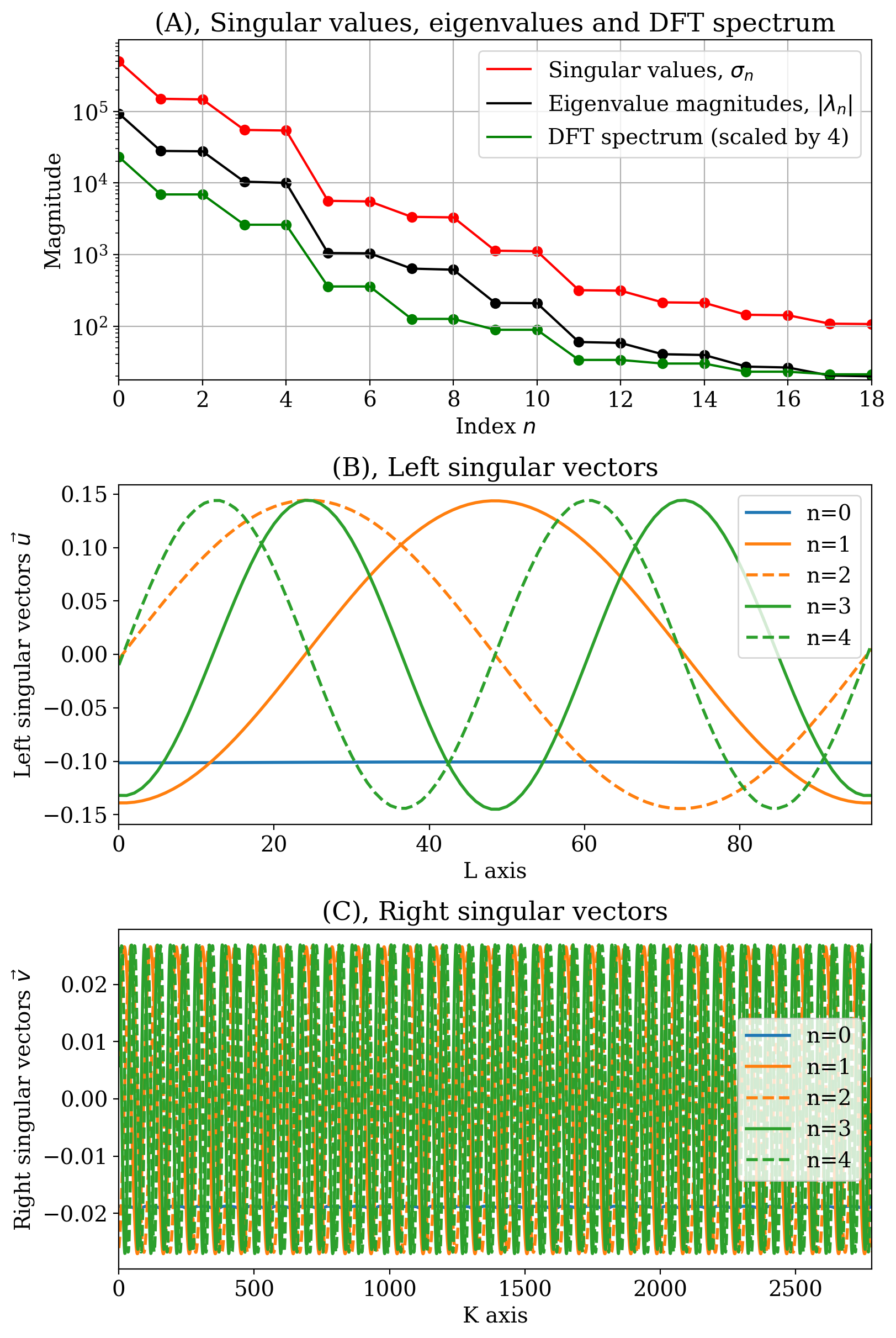}
    \caption{SVD of the trajectory matrix. Panel (A) shows singular values in a semilogarithmic scale. The same panel shows the eigenvalues of an $L \times L$ submatrix as well as the DFT spectrum of the periodic sequence. The DFT spectrum has been sorted according to descending magnitude and artificially scaled by a value of 4 to make it distinguishable from the eigenspectrum. For clarity, only the first 19 values are plotted. Panel (B) shows the first 5 left singular vectors while Panel (C) shows the first 5 right singular vectors. In both plots, the orthogonal (sine-cosine) vector pairs are plotted with the same colour but with different line-styles.}
    \label{fig:SVD_eigen_fft_compare}
\end{figure}

The trajectory  matrix  is then decomposed with SVD. The results from the SVD are shown in Fig.\ref{fig:SVD_eigen_fft_compare}. Independently, an $L \times L$ submatrix of the trajectory matrix (with $L=96$) is eigendecomposed. The eigenspectrum of this anti-circulant matrix is also given in Fig.\ref{fig:SVD_eigen_fft_compare}. Moreover, the DFT of the underlying periodic sequence of length $L=96$ is also computed and plotted in the same figure with the DFT values sorted in descending order. As the sequence is real, the DFT spectrum is Hermitian and the sorted values appear twice in the spectrum. 

Several observations can be made from Fig.\ref{fig:SVD_eigen_fft_compare}. The spectrum of singular values matches the spectrum of eigenvalue magnitudes exactly, except for a scaling (which is  $\sqrt{29}$). The spectrum of eigenvalue magnitudes matches exactly the DFT magnitude spectrum, thereby validating the results from Appendix.\ref{Ap:acirc_eigen}. The same scaling is also evident in the amplitudes of the right singular vectors. We also find that the singular vectors, except the $n=0$ component, are purely sinusoidal and occur in sine-cosine pairs as the calculations showed. We have also verified that the vector pairs are indeed orthogonal by calculating the inner product between such pairs. The singular vectors are also periodic in $L$ while the $n=0$ component is essentially the DC component of the data, similar to the zeroth component in Fourier transforms. 

\begin{figure}
\centering
	\includegraphics[width=\columnwidth]{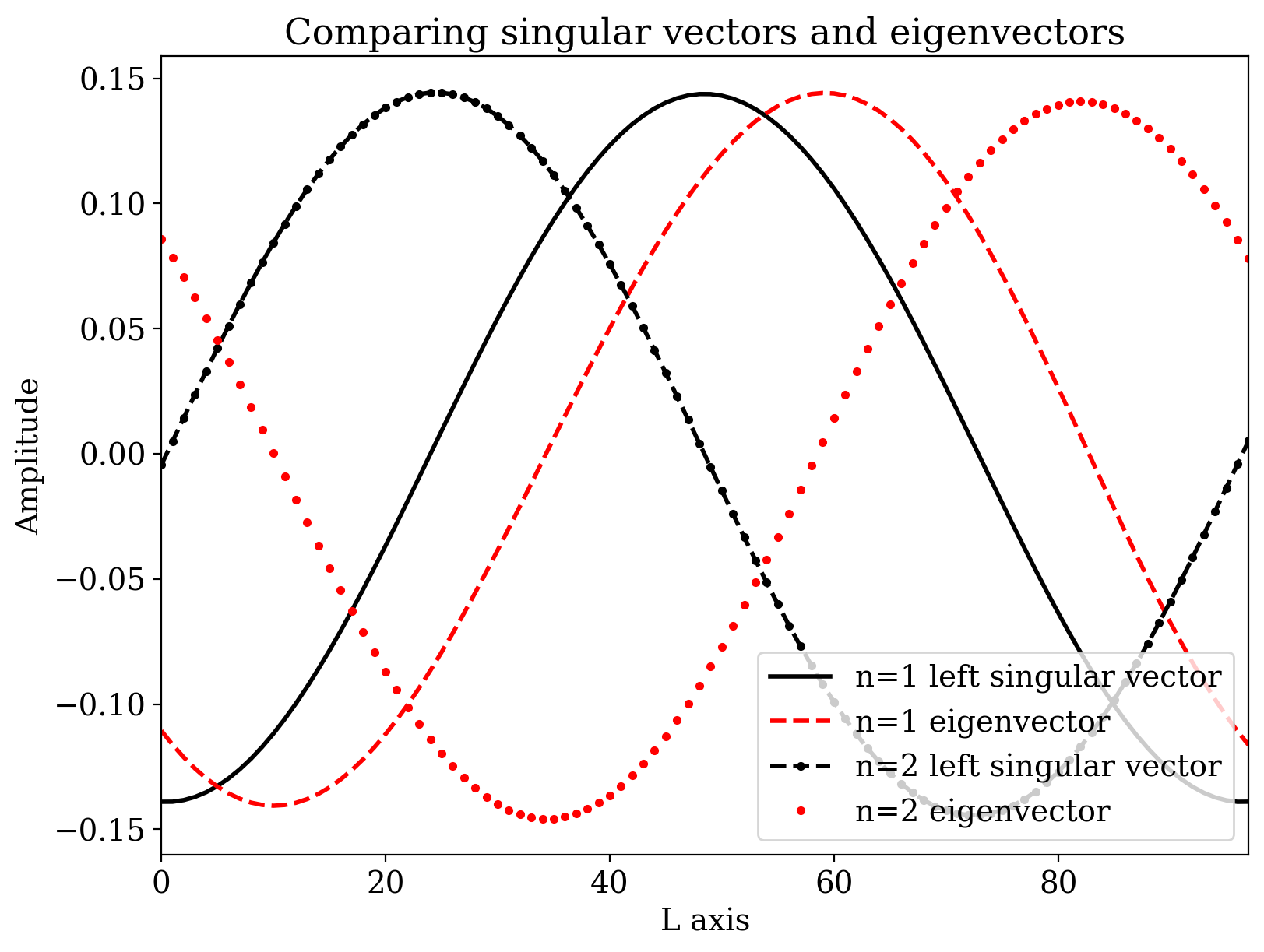}
    \caption{A comparison between the left singular vectors of the trajectory matrix and eigenvectors of a square submatrix. Only the vectors from the first orthogonal pair in each case are plotted.}
    \label{fig:SVD_eigen_vect_compare}
\end{figure}

In Fig.\ref{fig:SVD_eigen_vect_compare}, we compare the first pair of orthogonal left singular vectors with the corresponding pair of the eigenvectors. We find that the eigenvectors are exactly the same as the singular vectors, except for a sign reversal in one of the vectors in the pair. It has been verified that the corresponding eigenvalue carries a negative sign, thus it is inconsequential to the overall analysis, and we can safely consider the equivalence between SVD and eigendecomposition in Eq.\ref{Eq:SVD_eig_expansion} to be valid. 

\begin{figure}
\centering
	\includegraphics[width=\columnwidth]{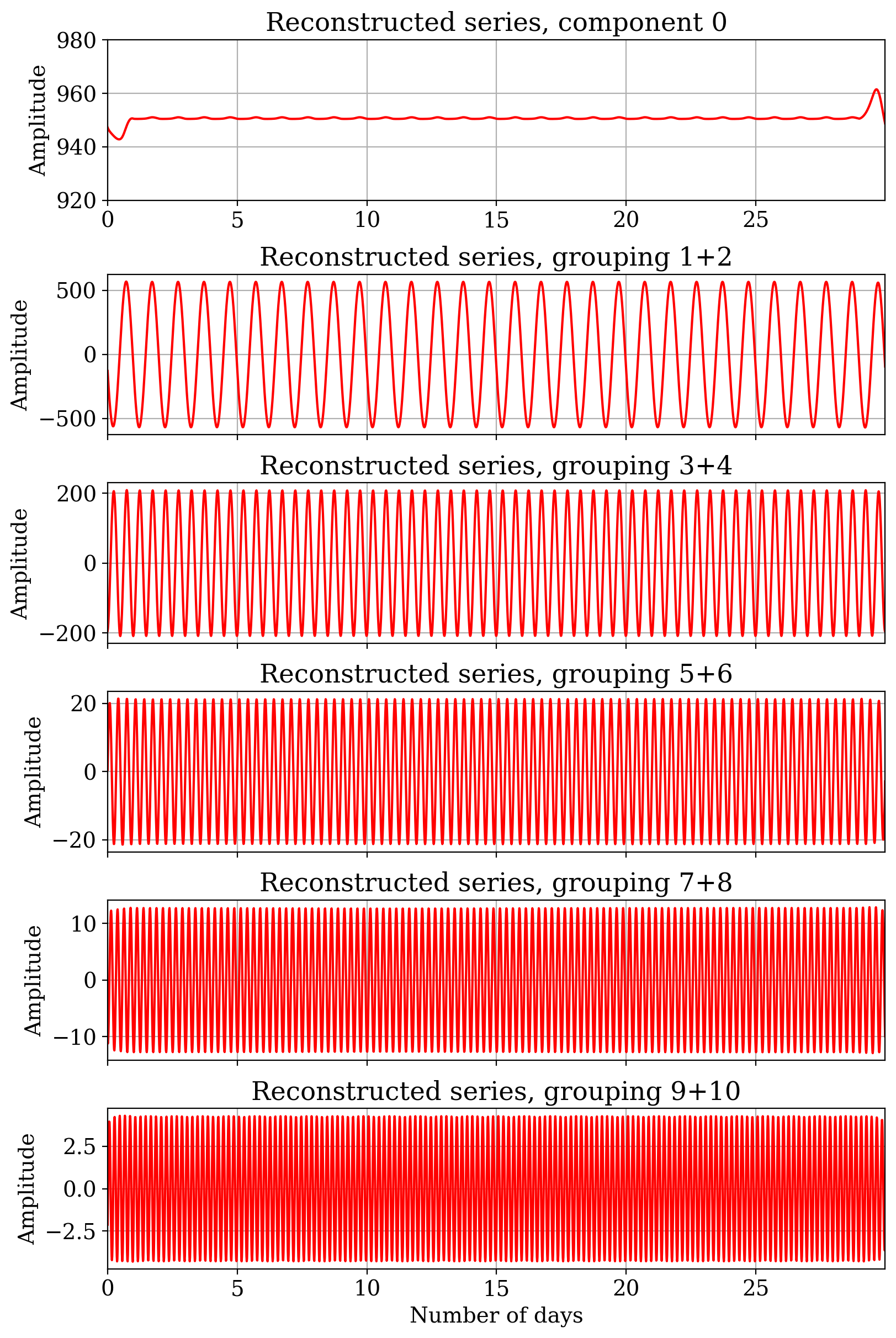}
    \caption{Reconstructed series from SSA of an ideal periodic series. The $0^{th}$ component and the succeeding 5 grouped grouped orthogonal pairs are given in this figure. Since the reconstructed series add up to the original time series, their $x$-axis is the number of days.}
    \label{fig:Recon_components}
\end{figure}

Thus, we find that the singular vectors obtained from SSA of a periodic sequence are purely sinusoidal with them occurring in orthogonal pairs. It may be also noted that while such orthogonal features have been noticed in the SSA literature (see for e.g. \cite{Ghil20021}), we mathematically demonstrated using simple arguments why such orthogonal sinusoidal pairs are to be expected in the SSA of an ideal time series. We now inspect the reconstructed series given in Fig.\ref{fig:Recon_components}. As we know that the singular vectors and corresponding values occur in pairs, we group the eigentriples into pairs (except the $0^{th}$ component) and apply the diagonal averaging. As can be seen, the reconstructed series are also purely sinusoidal since the vectors are sinusoidal.  Another important observation with Fig.\ref{fig:Recon_components} is the $0^{th}$ component, which is often called ``trend'' in SSA literature. For radiometric data this term corresponds to the sky-averaged component of the radio data and is the relevant observable for global 21--cm research. Since the origin of the $0^{th}$ reconstructed component can be traced back to $0^{th}$ eigenvector of the matrix $\bm{Q}$, we can see that the corresponding Fourier component has a frequency of zero. In other words for an ideal periodic time series, the $0^{th}$ component in SSA is related to the zero frequency or DC term in the Fourier spectrum, such that its singular value is the same as the Fourier DC term.  

In the above exercise, it is essential that the embedding dimension (window length) corresponds to exactly one sidereal day. If not, the decomposition will not lead to sinusoidal patterns of appropriate periodicity, and the various relations that we obtained between SVD, eigendecomposition and Fourier analysis become invalid. With this important result, we now proceed to introduce some nonidealities into the time series.

\subsection{SSA applied to periodic time series with time-varying gains}
The calculations and simulations so far assume a case where the radiometer system has been fully calibrated to yield data calibrated to a reference plane outside the Earth's ionosphere. Observational radiometric data have multiple nonidealties, the dominant ones are additive components such as receiver noise temperatures and mutliplicative factors that are often called gains. Both of them can be time-varying and therefore are required to be known to correct the data. We now incorporate them into our model. 

Assuming a simple system model for SITARA data (see Sec. 5.2 of \cite{2022PASA...39...18T}), the trajectory matrix of the measured data can be expressed as a Hadamard product, given in Eq.\ref{Eq:sys_eq}.
\begin{align}  
    \label{Eq:sys_eq}
    \bm{X'} &= (\bm{X_s + R})\odot \bm{G} 
\end{align}
where $\bm{X_s}$ is the sky temperature trajectory matrix, $\bm{R}$ is the receiver noise temperature trajectory matrix and $\bm{G}$ is the trajectory matrix of time-varying system gains. Since the low frequency radio sky is bright, the dominant time-varying component in the data is due to sky temperature multiplied with time-varying system gains. Therefore, we can apply a simplifying assumption that the receiver noise temperature is constant with time and absorb it into the ``ideal'' sky matrix, and the sum can be written as $\bm{X}$. Thus we write 
\begin{align}  
    \label{Eq:sys_eq_X}
    \bm{X'} &= \bm{X}\odot \bm{G}. 
\end{align}
While Eq.\ref{Eq:sys_eq} and \ref{Eq:sys_eq_X} may be written without resorting to a  matrix formulation, expressing them with trajectory matrices enables SSA.

Based on the summation of $\bm{X}$ from Eq.\ref{Eq:X_expansion} and Eq.\ref{Eq:sys_eq_X}, the following relations can be derived.
\begin{align}
\label{Eq:sys_eq_Tcirc}
\bm{X'} &= \sum_i \bm{X'_{i}}, \rm{and} \\ \nonumber 
\bm{X'} &= \Big(\sum_i \bm{X_{i}}\Big) \odot \bm{G} \\ \nonumber
      &= \sum_i (\bm{X_{i}}\odot \bm{G})
\end{align}
where we have used the distributive property of Hadamard products. Thus we have,
\begin{equation}
\label{Eq:X_i2Xp_i}
    \sum_i \bm{X'_{i}} = \sum_i (\bm{X_{i}}\odot \bm{G}).
\end{equation}
Though Eq.\ref{Eq:X_i2Xp_i} appears trivial, it reveals a  useful aspect of the decomposition.  Applying diagonal averaging to each $\bm{X'_{i}}$ leads to gain multiplied reconstructed series, as the Hadamard product is an element-by-element multiplication. However, we already know the ``true'' reconstructed series to be purely sinusoidal from Sec.\ref{sec:SSA_periodic}.  The multiplication of a pure sinusoid waveform by a signal of much lower frequency results in amplitude modulation of the sinusoid, a process that is called amplitude modulation (AM) in communication systems \citep{Haykin_Moher}. Therefore, we may treat the gain multiplied reconstructed series as a form of AM signals with the sinusoids acting analogous to ``carriers''; and using AM demodulation techniques to extract the temporal variations in gains.  

\begin{figure}
\centering
	\includegraphics[width=\columnwidth]{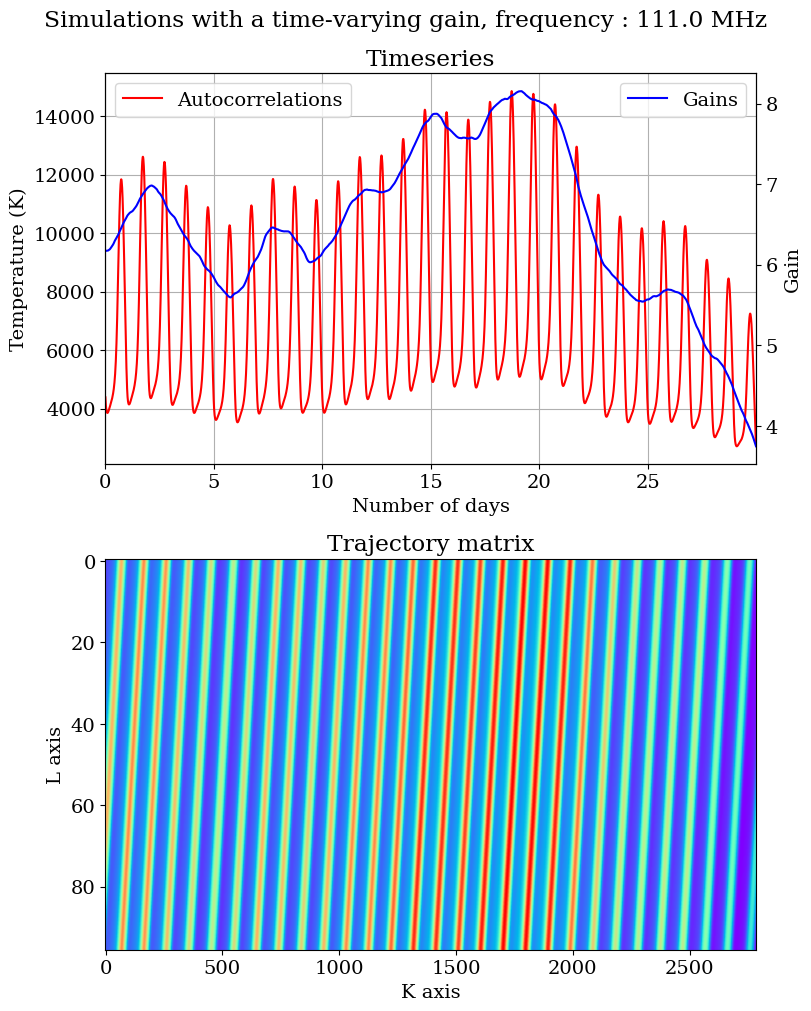}
    \caption{Simulated time series with gains at a frequency of 111~MHz and the corresponding trajectory matrix}
    \label{fig:Simu_P_111_trajectory_w_gain}
\end{figure}

\begin{figure}
\centering
	\includegraphics[width=\columnwidth]{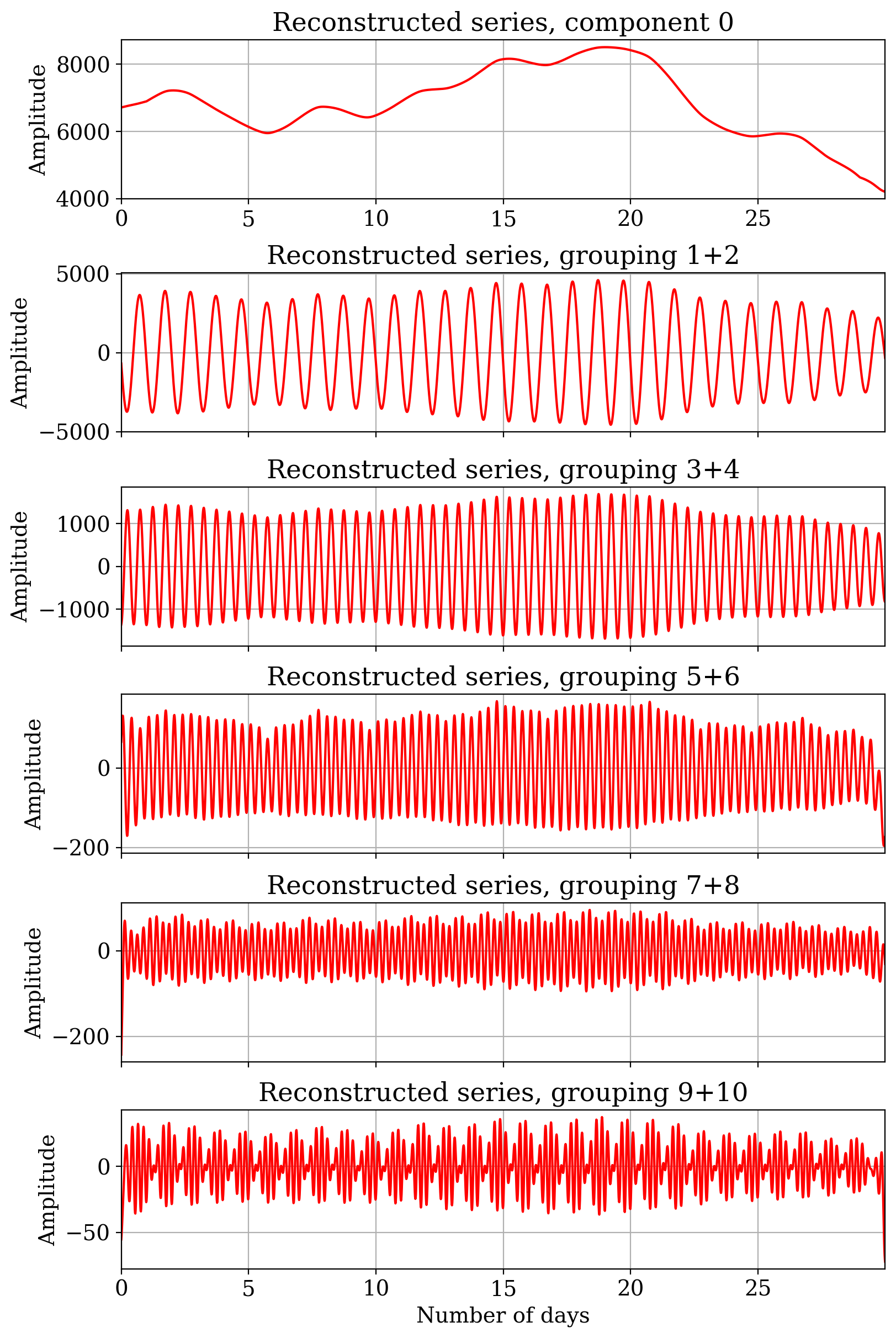}
    \caption{Reconstructed series from SSA of a simulated series with time-varying gains. The $0^{th}$ series and the succeeding 5 components from grouping orthogonal pairs are given in this figure.}
    \label{fig:Recon_components_wgains}
\end{figure}

Before proceeding further, it is instructive to verify the above calculations with simulations. For this, we begin with the basic time series from Sec.\ref{sec:SSA_periodic}. To this series, we add a constant of $100~K$ as the receiver noise temperature and simulated radiometric noise (integration time of 15 minutes and bandwidth of 61~kHz) for the overall system temperature.  The gain profile is modelled as a random-walk, which is generated as a cumulative sum of Gaussian white noise. The resulting profile is subsequently smoothed with box-car averaging to suppress variations of time cadence less than one sidereal day. The gains are also normalised to avoid negative excursions. The time series is then multiplied with the simulated gains and the resulting series and the associated trajectory matrix are shown in Fig.\ref{fig:Simu_P_111_trajectory_w_gain}.  It may be noted that the dimensions of the matrices and time series have all been kept the same.
The trajectory matrix is then decomposed with SVD, and the eigentriples grouped in the same manner as in Sec.\ref{sec:SSA_periodic} and diagonally averaged. It is easy to notice that the resulting reconstructed series shown in Fig.\ref{fig:Recon_components_wgains} are indeed amplitude modulated sinusoids.

\begin{figure}
\centering
	\includegraphics[width=\columnwidth]{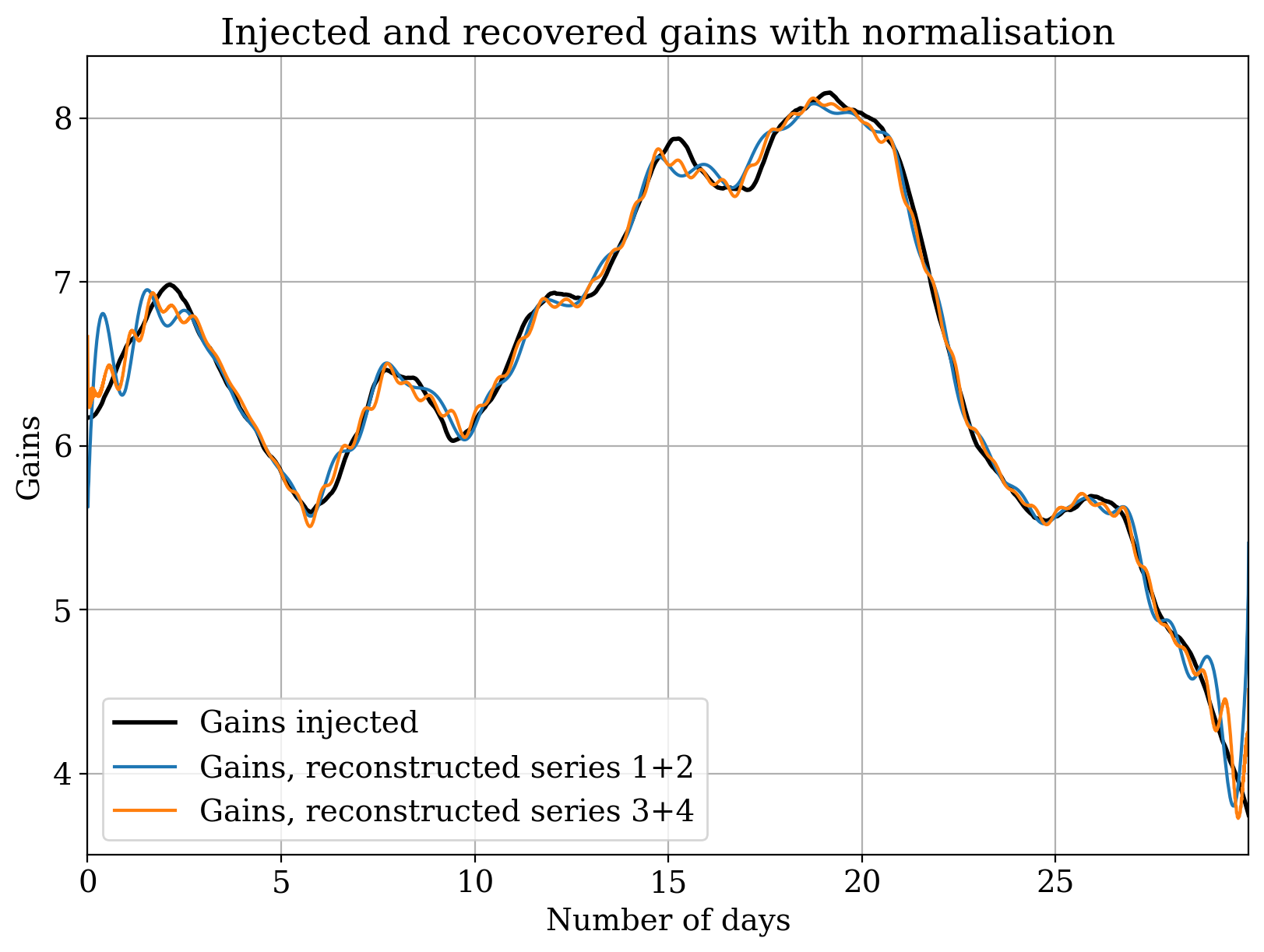}
    \caption{Injected and recovered gains. The recovered gains have been normalised according Eq.\ref{Eq:x_gain_reco}. The recovered gains from periodic components resemble the injected gains.}
    \label{fig:Gain_recovery_sims_wnormalisation}
\end{figure}

Several algorithms exist to demodulate such AM signals to obtain their envelopes. Here we use an algorithm that is mathematically simple to interpret. The first step in this algorithm is converting a real valued signal into an analytic signal, which can be accomplished with Hilbert transforms. Taking the magnitude of this analytic signal yields the modulation envelope. We apply this procedure to the series given in Fig.\ref{fig:Recon_components_wgains} . The resulting gain profiles are plotted in Fig.\ref{fig:Gain_recovery_sims_wnormalisation}, in which the curves are also normalised according to the method described in Sec.\ref{subsec:SSA_cal}. 

Fig.\ref{fig:Gain_recovery_sims_wnormalisation} validates the interpretation of Eq.\ref{Eq:X_i2Xp_i} that the reconstructed series have identical gain templates that can be recovered with amplitude demodulation. Indeed, a major outcome of this exercise is that the gain template for the DC component is identical to those of the periodic components. Therefore the time-variations in the DC term may be corrected for by using the gain templates from periodic components. This leads us to a novel calibration strategy for global 21--cm radiometers that can correct long-term drift and related systematics.

\subsection{On the use of SSA to aid calibration}
\label{subsec:SSA_cal}
 The recovered gains require normalisation to enable a comparison with the injected gains and make them useful for calibration.   
We begin with Eq.\ref{Eq:X_i2Xp_i} and incorporate Eq.\ref{Eq:X_expansion} as shown in Eq.\ref{Eq:X_svd_expansion}, where we also include the eigentriple grouping. The indices have also been modified for convenience.
\begin{align}
    \label{Eq:X_svd_expansion}
    \bm{X'_i} &= \bm{G} \odot \sum_{j=0}^{1}\sigma_{i+j}   (\vec{u_{i+j}} \otimes \vec{v_{i+j}}), i=1,3... \\ \nonumber
   \bm{X'_i} &= \sigma_{i}\bm{G} \odot \sum_{j=0}^{1}   (\vec{u_{i+j}} \otimes \vec{v_{i+j}}), i=1,3... 
\end{align}
where we used the property that singular values are identical for the eigentriples in a grouped pair. The vectors $\vec{u}$ and $\vec{v}$ are from unitary matrices and therefore their individual inner products equal to unity. However, for subsequent analysis, these vectors need to be normalised to have unity \textit{amplitude} when diagonally averaged after grouping. The normalisation can be achieved with the following operations, where the sinusoidal terms in the brackets have been normalised to have unity amplitude.
\begin{align}
    \label{Eq:u_v_norm}
    \vec{u_i} &= \frac{1}{\sqrt{L/2}}\big(\vec{u_i}\sqrt{L/2}\big)\\ \nonumber
    \vec{v_i} &= \frac{1}{\sqrt{K/2}}\big(\vec{v_i}\sqrt{K/2}\big)
\end{align}
as the lengths of the vectors $\vec{u}$ and $\vec{v}$ are $L$ and $K$ respectively. 
The grouped and diagonally averaged reconstructed series is given as
\begin{equation}
    \label{Eq:x_gains_expansion}
    \widetilde{x_i}[n] = \frac{\sigma_i} {\sqrt{L/2} \sqrt{K/2}}~ g[n] c_i[n].
\end{equation}
where $g[n]$ is the gain series and $c_i[n]$ is the ``carrier'' sinusoid. The normalisation in Eq.\ref{Eq:u_v_norm} ensures unity amplitude for $c_i[n]$.  Upon demodulation of a series $\widetilde{x_i}[n]$, the carrier is removed and a modulation envelope $g_{i, t}[n]$ is obtained as 
\begin{equation}
    \label{Eq:rec_gain_template}
    g_{i, t}[n] =  \frac{\sigma_i} {\sqrt{L/2} \sqrt{K/2}}~ g'_{i}[n].
\end{equation}
 From these envelopes, gain templates $g'_{i}[n]$ can be recovered as 
\begin{equation}
    \label{Eq:x_gain_reco}
    g'_{i}[n] =  g_{i, t}[n]\frac{\sqrt{L/2} \sqrt{K/2}}{\sigma_i}.
\end{equation}
For Eq.\ref{Eq:x_gain_reco} to be useful for calibration, the only auxiliary information required is the singular value ${\sigma_i}$. This is the singular value of the component with the same periodicity of that we expect from the true sky, and can be obtained from the SSA of the simulated ideal sky time series from Sec.\ref{sec:SSA_periodic}. Using those singular values, we obtain the recovered and appropriately normalised gains that are plotted in Fig.\ref{fig:Gain_recovery_sims_wnormalisation}. 

\begin{figure}
\centering
	\includegraphics[width=\columnwidth]{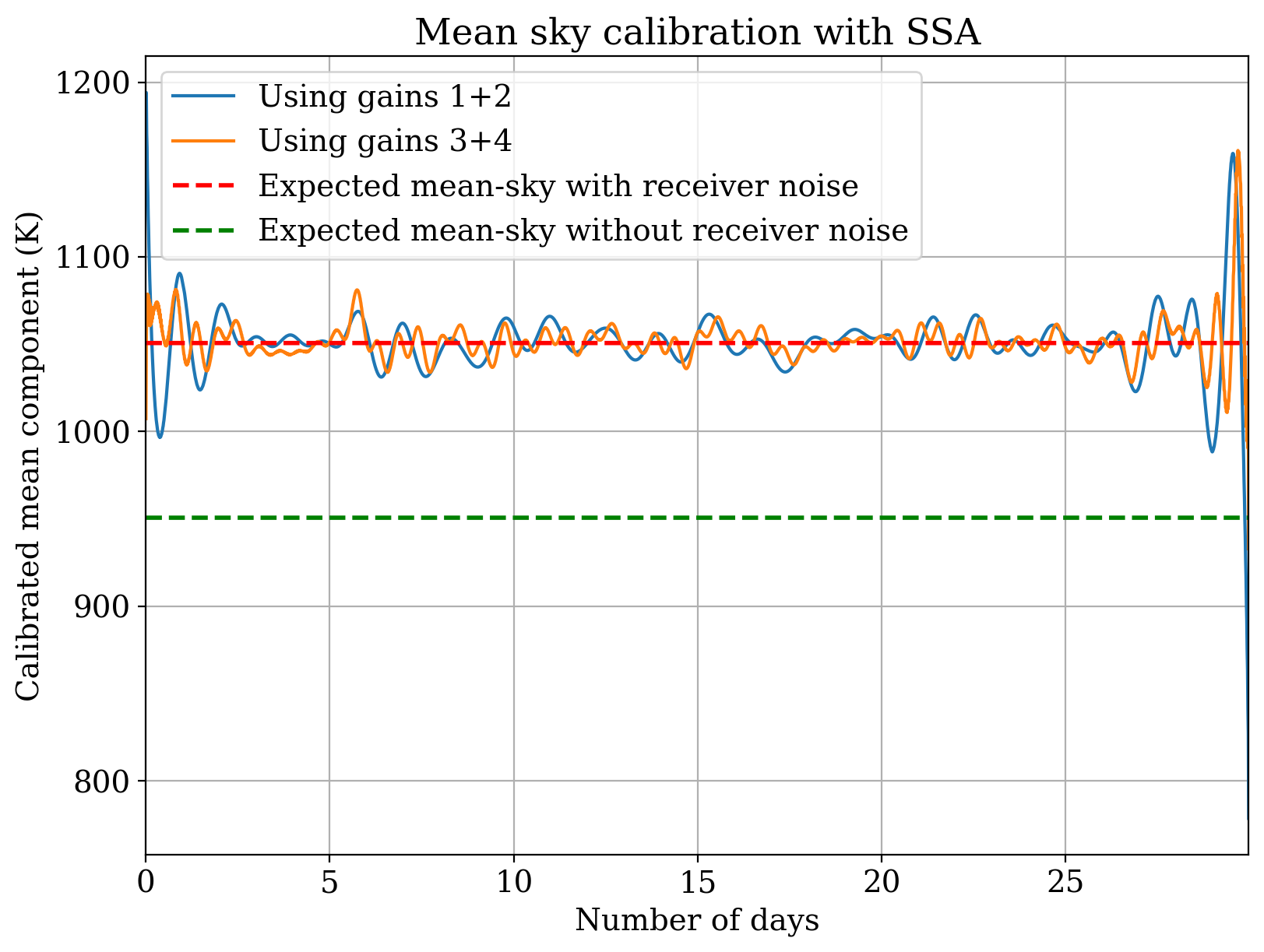}
    \caption{Calibrating the mean sky component using recovered gains from periodic components. For comparison, the expected mean levels with and without the 100~K receiver noise temperatures are also plotted. }
    \label{fig:Mean_sky_cal_sim}
\end{figure}

We are now in a position to apply those gains to the $0^{th}$ reconstructed series to achieve calibration and compare the calibration with the expected levels. The results are given in Fig.\ref{fig:Mean_sky_cal_sim}.
We find that the mean-sky component has been calibrated solely using the periodic component(s) of the sky with the aid of SSA. It has to be noted that the prescription provided here to calibrate the mean-sky relies only on the singular values of the simulated periodic components. Even without using the singular values, and only using the data, it is possible to achieve relative gain calibration, while applying gains with appropriate singular values to the mean-sky component establishes its proper brightness temperature scale.

Let us now try to obtain an intuitive appreciation of this calibration. The trend component obtained from SSA is essentially a box-car averaging of time series data. If the window used for averaging encompasses exactly one sidereal day, the periodic variations caused by the Galaxy transiting through the beam are averaged out. If the system were perfect the result would be a constant temperature, which is the sum of mean-sky and receiver noise temperatures. Thus the trend that we observe in the non-ideal system may be modelled by a constant value multiplied with the time-varying gains. 

The gain variations that we have studied so far have smooth evolution over time with no periodicity. However, for real radio telescope systems that do not have temperature regulation, gain variations with local temperature are expected - details given in Sec.\ref{sec:sitara_SSA}. Indeed, the rising and setting of the Sun can heat and cool the components in the analog signal chain and induce diurnal variations in the gain that may be correlated with the sky-drift. Yet another potential origin for diurnal patterns in gains is the ionosphere. Therefore it is imperative to study such a scenario, which we perform next.

\subsection{SSA with diurnal gain variations}
\label{subsec:diu_gain_sims}
\begin{figure}
\centering
	\includegraphics[width=\columnwidth]{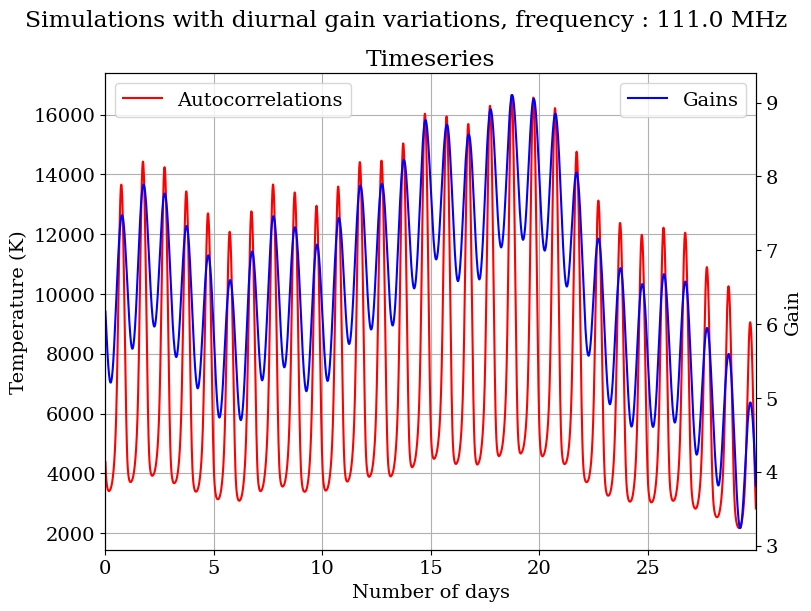}
    \caption{Simulated radiometer data with gains that have a smoothly varying component and a diurnal component.}
    \label{fig:Simu_P_111_series_w_gain_perio}
\end{figure}

\begin{figure}
\centering
	\includegraphics[width=\columnwidth]{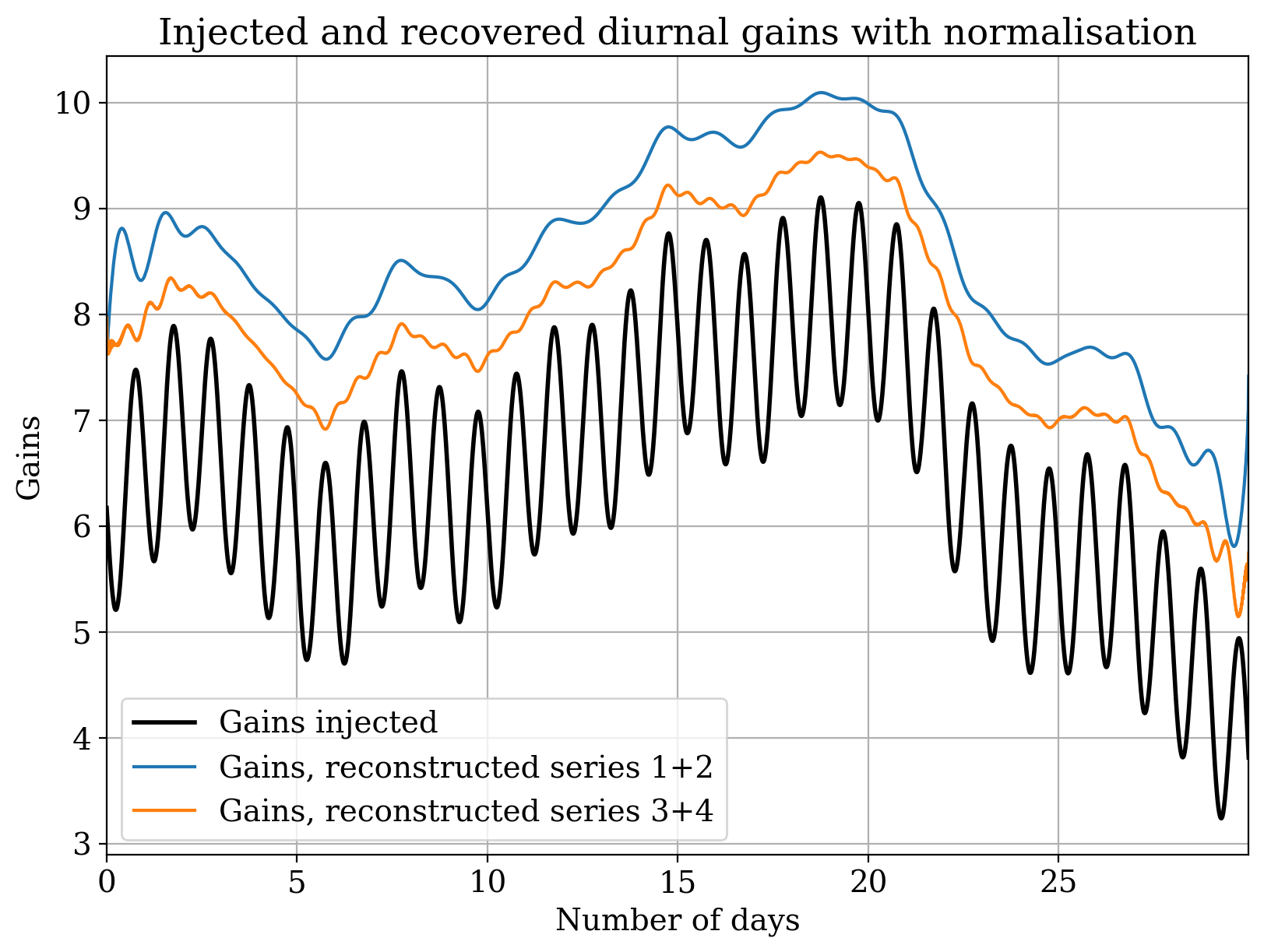}
    \caption{Injected and SSA recovered gains when the gains have a diurnal component. Appropriate normalisation has been applied.}
    \label{fig:Gain_recovery_sims_perio_wnormalisation}
\end{figure}

\begin{figure}
\centering
	\includegraphics[width=\columnwidth]{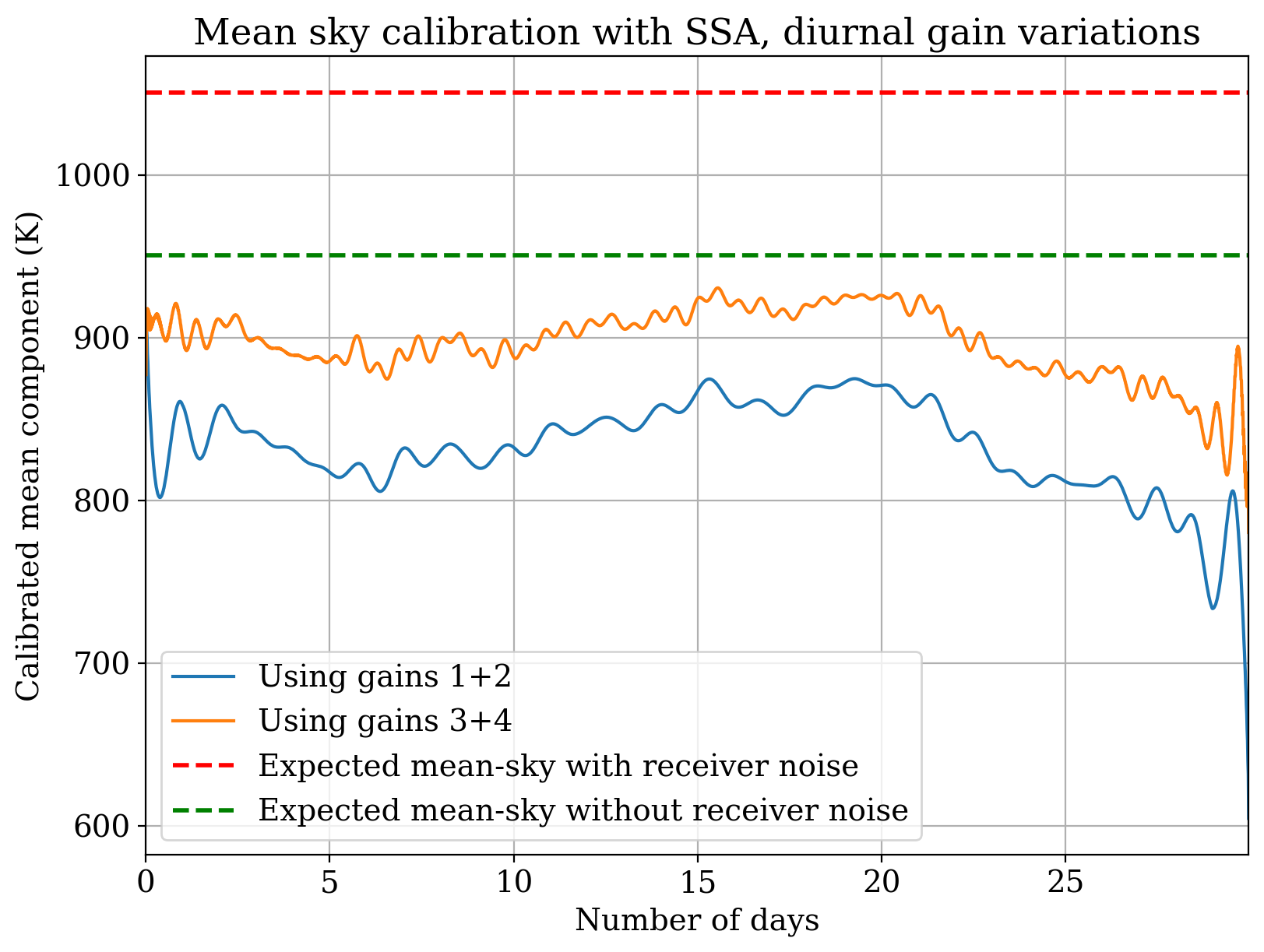}
    \caption{Application of recovered gains from periodic components to calibrate the mean-sky term, when the injected gains have a diurnal component.}
    \label{fig:Mean_sky_cal_sim_perio}
\end{figure}

To the gains simulated in the previous section, we add a sinusoidal diurnal gain component. For a time series that spans a month, the diurnal gain variations are expected to show a strong correlation with Galaxy transit. For longer time series, the difference between sidereal time and civil time reduces this correlation. The modified gains and the resulting time series are shown in Fig.\ref{fig:Simu_P_111_series_w_gain_perio}. 
We perform SSA in the exact same manner as in the previous case, and recover normalised gains, as shown in Fig.\ref{fig:Gain_recovery_sims_perio_wnormalisation}.
It is interesting to note that the correlation between gains and sky patterns reduces the separability of the two, and the recovered gains from the reconstructed series differ from one another. We attempt a calibration of the mean component with these gains and the results are given in Fig.\ref{fig:Mean_sky_cal_sim_perio}.
% (for example, \texttt{scipy.signal.hilbert})
The differences in the recovered gains in Fig.\ref{fig:Gain_recovery_sims_perio_wnormalisation} as well as the calibrated temperatures in Fig.\ref{fig:Mean_sky_cal_sim_perio} demonstrate a fundamental limitation of the SSA technique in calibrating data when the gain variations are correlated with sky patterns. Lacking auxiliary information  related to the cause of such gain variations, for example temperature measurements, \color {black} the calibration technique fails and yields erroneous results as demonstrated. On the other hand, a disagreement between the recovered gains can be used as a practical indicator of gain variations that have frequencies coinciding with the periodicity induced by sky drift, thus pointing to a necessity for subsidiary information.  In other words, differences in recovered gains between the groupings indicate presence of correlations between the gains and sky drift. It may be also noted that no \textit{a priori} information is required to infer such a correlation. Having demonstrated the various cases of SSA, we now apply these techniques to data from SITARA observations.  

\section{SSA applied to SITARA data}
\label{sec:sitara_SSA}
\subsection{Data preparation}
We use a time series of SITARA spectral data from the concatenated 30 days of June 2021 data. The individual data files in \texttt{miriad} format are converted to \texttt{hdf5} format and concatenated using custom tools written in \texttt{python}. From the concatenated dataset, auto-correlation data for a frequency of 111~MHz are extracted, as it is a frequency with relatively low RFI occupancy. The major source of RFI observed during this period was Solar radio bursts that are transient in nature. This raw data time series as a function of Julian Date (JD) is shown in Fig.\ref{fig:raw_data_111MHz}. 

The raw SITARA data spanning a month have about 749,000 points, which makes calculations difficult. Therefore, the data are binned into the same number of bins (2879) as done in the simulations, giving one sample per 15 minutes or 96 samples per sidereal day. Once again, in performing binning it is important to use sidereal time - and not Julian time - to ensure an equal number of samples for each cycle of sky-drift. The data are then embedded into a trajectory matrix and SSA is applied in the same manner as before. The eigentriples are grouped and reconstructed series formed. 

\begin{figure}
\centering
	\includegraphics[width=\columnwidth]{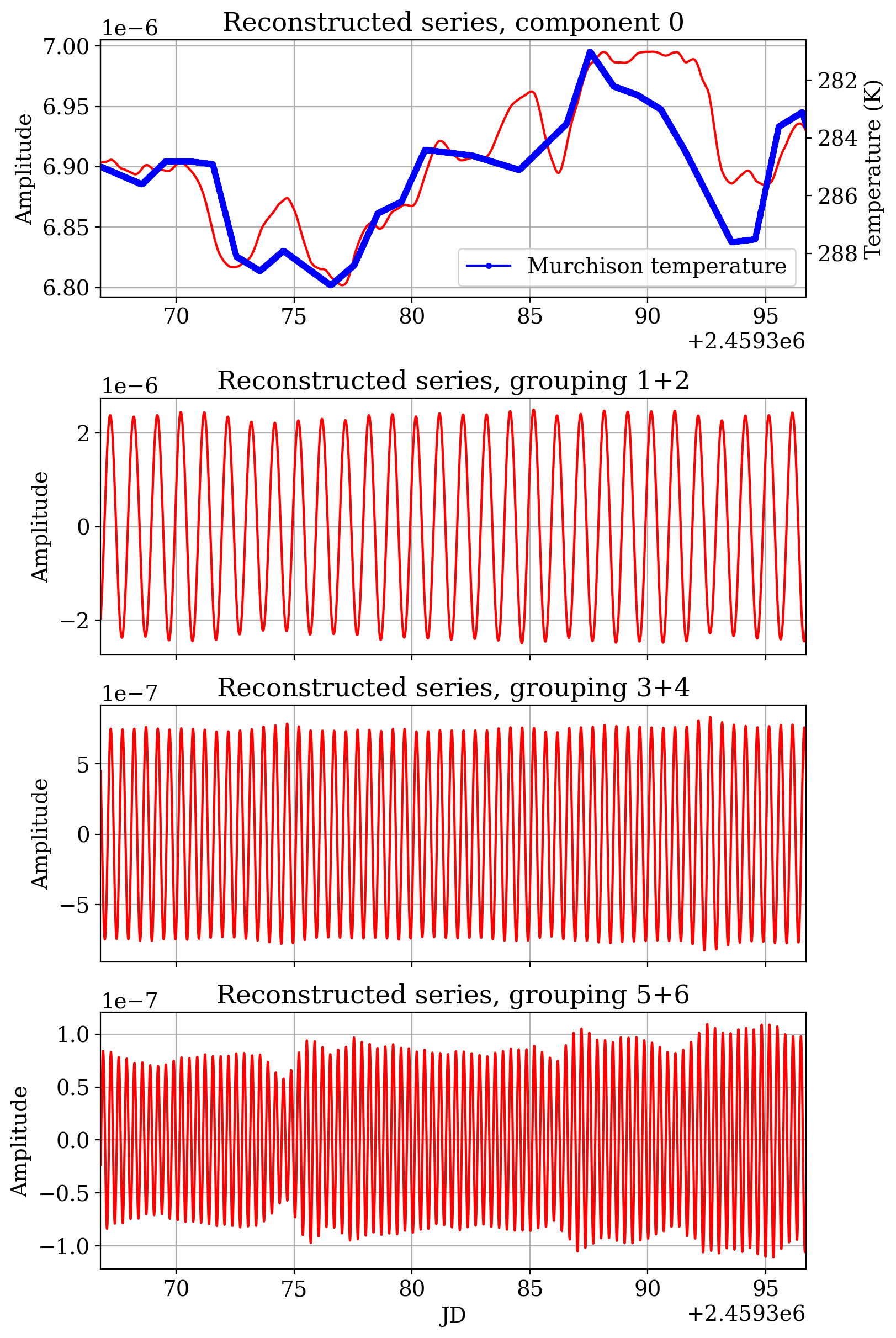}
    \caption{Reconstructed series from SITARA data SSA. A plot of the physical temperature recorded within the Murchison Shire is included in the trend plot to show the anti-correlation between them.}
    \label{fig:SITARA_Recon_comp_wgains_MROtemp}
\end{figure}

Fig.\ref{fig:SITARA_Recon_comp_wgains_MROtemp} shows first 4 of the reconstructed series. The trend plot also has a low time cadence record of the temperature recorded within the Murchison shire, which is available from NCEI-NOAA  \footnote{\url{https://www.ncei.noaa.gov/access/search/index}}. The temperature scale has been inverted, as the trend shows an anti-correlation with temperature. Further, the radio data and temperature data fall into two different families with different sampling, and therefore the temperature data have been interpolated to the same Julian day bins as the radio data. The Pearson correlation coefficient computed between the trend and the temperature data is -0.83, which denotes a strong anti-correlation between the trend and temperature. 

Apart from component ageing effects, there are two major reasons for such temperature-induced variations in radiometric data.
\begin{enumerate}
    \item The noise temperatures of active devices used in amplifiers increase with physical temperature\footnote{It may be noted that this is one of the reasons for cryogenic cooling of radio telescope front-ends to achieve a low overall system temperature.}. The same holds true for passive components such as attenuators. If this is the cause of drift, the pattern so obtained is expected to be \textit{correlated} with physical temperature.
    \item The gain of amplifiers reduces as temperature is increased. In this case, the trend pattern and physical temperature would be \textit{anti-correlated}.
\end{enumerate}

It is therefore evident from Fig.\ref{fig:SITARA_Recon_comp_wgains_MROtemp} that the major contributor to the trend is temperature-induced gain variations. The receiver noise would inevitably vary as a function of physical temperature, however when the overall system temperature is sky-dominated the impact of this would be secondary to gain variations. This informs our choice of calibration model given in Sec.\ref{subsec:SSA_cal}, where we assume a constant receiver noise temperature and a time-varying gain. 

\begin{figure}
\centering
	\includegraphics[width=\columnwidth]{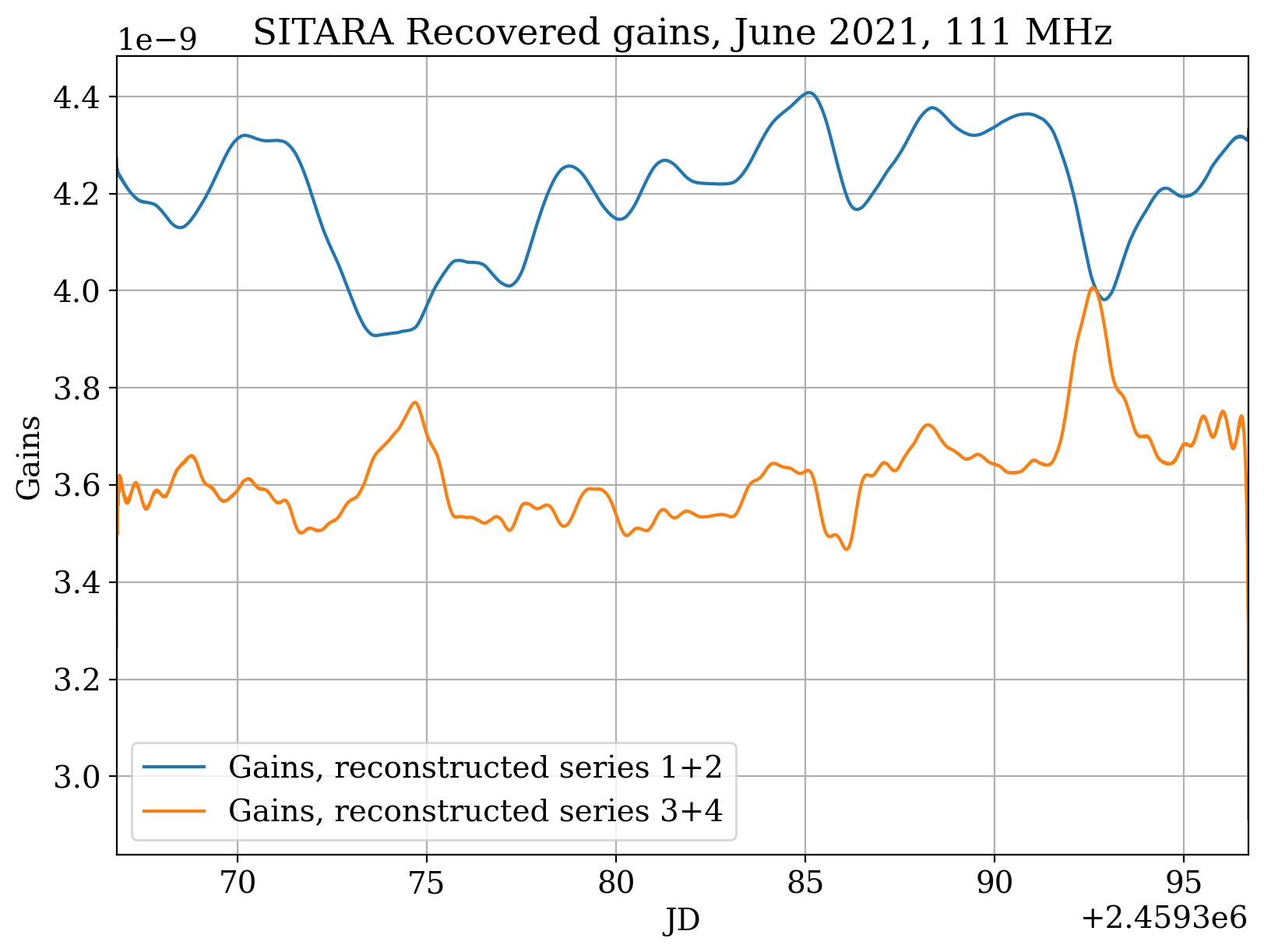}
    \caption{Recovered gains from SITARA data, June 2021. The gains differ across reconstructed series, pointing to potential diurnal gain variations.}
    \label{fig:SITARA_Gain_recovery_wnormal}
\end{figure}

Subsequently, the gain patterns are recovered from the reconstructed series with the demodulation technique. These recovered gains for the first two periodic components are shown in Fig.\ref{fig:SITARA_Gain_recovery_wnormal}. As can be seen, the normalised gains are different between the series, thus pointing to potential diurnal gain variations. Therefore, we refrain from applying these gains to the mean component to establish its brightness temperature scale.

\section{Discussion}
In this paper we developed the mathematical framework for SSA of a radiometric time series and demonstrated its application in analysing radiometric time series data from a radio telescope at a frequency with relatively low RFI. A major outcome of this work is a novel method to calibrate the mean-sky component of radio data using the periodic component of sky-drift patterns.  Calibration using differential measurements of the sky to obtain the system gain has been successfully utilised in different low frequency experiments \citep{https://doi.org/10.1029/2003RS003016, Singh_2017, 2022PASA...39...18T}. However, those applications were limited, as the calibration was performed on a day-to-day basis assuming a constant value for the system gain for a specific day. With SSA, the periodic nature of data is leveraged to obtain gains that can vary with time, provided they do not a diurnal component. Seen in this light, SSA can be viewed as an extension of the aforesaid technique.  

Let us examine the implications of the gain recovery in Fig.\ref{fig:Gain_recovery_sims_wnormalisation}. The gain patterns that we recovered are purely from the periodic components of the sky and the singular values of simulated periodic components. The procedure does not need simulations that include the mean-sky or the \textit{zero-point} of the sky maps. Indeed, zero-point errors have been noticed in sky maps, since these maps are often products of combination of sky surveys conducted with different telescopes. Disagreements exist between estimates of the zero-points of sky maps across experiments, for example, \cite{2021ApJ...908..145M} report a different zero-point offset for the 150~MHz sky map \citep{1970AuJPA..16....1L} compared to the measurements of \cite{2015ApJ...801..138P}.  Therefore, the methods outlined in this paper can be applied to experiments where zero-point levels have to be accurately known to verify that the gains do not have variations that can confuse sky signals. However, to get accurate estimates of system gains the input sky maps need to get an accurate temperature scale even though zero-points errors are acceptable. An associated caveat is that while SSA can provide multiplicative system gains, the additive receiver noise temperature is practically indistinguishable from the mean-sky component. Therefore, experiments requiring accurate measurements of zero-level still need additional calibration to estimate the receiver noise temperature. Nonetheless, as demonstrated in this paper, SSA can be utilised as a diagnostic tool for such experiments.  
\subsection{Caveats and future work}
Since this is the first work exploring SSA for radiometric data analysis and calibration, we have kept the analysis simple. Specifically, low frequency radio data can have RFI and we ignored the flagging that has been performed. Flagging considerations lead to gaps in a time series dataset and may necessitate data in-painting or interpolation. Such explorations will be taken up in future work. 

We have developed the mathematical background and demonstrated an application of SSA for auto-correlations at a single frequency, while SITARA has nearly two thousand usable frequency channels in auto-correlations and cross-correlations each. To use all the frequency channels and to consider complex cross-correlations, the mathematical framework developed in this paper has to be expanded. For a proper treatment of multiple frequency channels, multivariate SSA techniques will be explored in future work. 

While we demonstrated via simulations a calibration technique to remove gain variations that evolve smoothly with time, the application of it to SITARA data is hampered by diurnal gain variations. This shows the necessity to maintain temperature regulation or inclusion of thermometers at points in the signal chain to track such gain variations - this will be incorporated in future revisions of the SITARA system. Additionally, the data in frequency channels below 50~MHz, where receiver temperature dominates over sky temperature, show diurnal periodicity arising from such gain variations which may be exploited as a template to improve calibration.  

\subsection{A potential application of SSA for space-based 21--cm experiments}
As we demonstrated, a major limitation with the SSA technique for calibration is the confusion between diurnal gain variations and sky drift, as both have approximately the same periodicity. If the periodicities can be made to differ, the calibration can be improved significantly. Specifically, if the sky-drift can be made faster than the gain variations, substantially better calibration can be expected. While it is difficult for ground-based experiments to introduce such a separation, a radiometer payload on a spin-stabilized satellite can have beams that rapidly scan the sky, thereby increasing the sky-drift rate. For example, a broadband dipole antenna could be placed on a satellite spinning such that the nulls sweep the Galactic plane at a rate much faster than any gain variations, including the flicker noise of the electronic systems. Spinning a satellite at a rate faster than than the $1/f$ knee frequency of the radiometric system, to reduce the deleterious impact of gain fluctuations on the images, has been employed in CMB missions such as Planck \citep{2010A&A...520A...4B}. As there are a few projects proposed, planned or launched targeting the global 21--cm signal from a satellite platform such as DARE \citep{2012AdSpR..49..433B}, Longjiang/Chang'e-4 \citep{2018P&SS..162..207J}, DSL \citep{2021RSPTA.37990566C}, PRATUSH etc., we opine that SSA would be an ideal tool for analysis and calibration of time series data from such radiometers in space. 

\section{Conclusions}
In this paper, we introduced singular spectrum analysis (SSA) as a powerful tool to study radiometric data. We showed the deep connections between the SSA techniques and Fourier transforms and leverage that to obtain long-term gain evolution from radio data. A novel technique to calibrate 21--cm experiments using periodicity in the sky drift patterns has been proposed and simulated. The limitations of that technique in the presence of diurnal variations that can confuse with sky-drift is studied. Upon application of SSA to SITARA data, we find that the obtained decomposition has the features as expected, with the trend showing strong anti-correlation with temperature. However, the gains obtained point to diurnal variations as a limiting factor in using SSA for gain calibration of SITARA data.

\section*{Acknowledgements}

This  work  is  funded  through CT’s  ARC  Future Fellowship, FT180100321 - ”Unveiling the first billion years:  enabling Epoch of Reionisation science”.
This work makes use of the Murchison Radio-astronomy Observatory, operated by CSIRO. We acknowledge the Wajarri Yamatji people as the traditional owners of the Observatory site. JNT acknowledges Dr. Adrian T. Sutinjo for his Linear Algebra over Lunch lectures. This work uses the following python packages and we would like to thank the authors and maintainers of these packages :- \texttt{numpy }\citep{2020Natur.585..357H}, \texttt{scipy }\citep{2020NatMe..17..261V}, \texttt{healpy }\citep{2019JOSS....4.1298Z}, \texttt{astropy }\citep{astropy:2013, astropy:2018}, \texttt{aipy }\citep{2016ascl.soft09012P}, \texttt{matplotlib }\citep{Hunter:2007}, \texttt{ephem }\citep{2011ascl.soft12014R}, \texttt{pygdsm }\citep{2016ascl.soft03013P} and \texttt{h5py} (\url{https://www.h5py.org/}).

%%%%%%%%%%%%%%%%%%%%%%%%%%%%%%%%%%%%%%%%%%%%%%%%%%
\section*{Data Availability}

The data used in this work will be made available based on reasonable requests.



\bibliographystyle{mnras}
\bibliography{SITARA_SSA}

%%%%%%%%%%%%%%%%%%%%%%%%%%%%%%%%%%%%%%%%%%%%%%%%%%

%%%%%%%%%%%%%%%%% APPENDICES %%%%%%%%%%%%%%%%%%%%%

\appendix

\section{Eigendecomposition of anti-circulant matrices}
\label{Ap:acirc_eigen}
The relation between circulant matrices and circular convolution as well as the relation between the eigendecomposition of such matrices and Discrete Fourier Transforms (DFT) are well known - see for e.g. \cite{CIT-006} for an approachable introduction to this relation - and put to use in several signal processing applications. However, the matrices of relevance for this paper are anti-circulant and therefore we provide some useful relations between those matrices and DFT.

Consider a real valued sequence $x[n], n=0,1... N-1$. This sequence can be used to construct an $N \times N$ real valued anti-circulant matrix as given in Eq.\ref{Eq:x_acirc}.
\begin{equation}
\label{Eq:x_acirc}
\bm{X_{ac}} = 
\begin{bmatrix}
x[0] & x[1] &...... & x[N-1] \\
x[1] & x[2] &...... & x[0] \\
x[2] & x[3] &...... & x[1] \\
..  & ..  &...... & .. \\
x[N-1] & x[0] &...... & x[N-2]\\
\end{bmatrix}\\
\end{equation}
The eigendecomposition of $\bm{X_{ac}}$ can be written as 
\begin{equation}
    \bm{X_{ac}} = \bm{Q\Lambda Q^T}
\end{equation}
where 
\begin{equation}
    \bm{X_{ac}}~y_n = \lambda_n y_n
\end{equation}
where $\lambda_n$ are the eigenvalues and $y_n$ corresponding eigenvectors with $n$ being the index.
The following set of equations are obtained from eigendecomposition.

\begin{align}
    x[0]y_n[0] + .... + x[N-1]y_n[N-1] &= \lambda_n y_n[0] \\ \nonumber
    x[1]y_n[0] + .... + x[0]y_n[N-1] &= \lambda_n y_n[1] \\ \nonumber
    ................................................... \\ \nonumber
    x[N-1]y_n[0] + .... + x[N-2]y_n[N-1] &= \lambda_n y_n[N-1]
\end{align}
which can be rearranged into,
\begin{align}
    x[0]y_n[0] + .... + x[N-1]y_n[N-1] &= \lambda_n y_n[0] \\ \nonumber
    x[0]y_n[N-1] + .... + x[N-1]y_n[N-2] &= \lambda_n y_n[1] \\ \nonumber
    ................................................... \\ \nonumber
    x[0]y_n[1] + .... + x[N-1]y_n[0]  &= \lambda_n y_n[N-1]
\end{align}

The above set of equations can be recast into a summation equation as given in Eq.\ref{Eq:ac_summation}.
\begin{align}
\label{Eq:ac_summation}
\sum_{k=0}^{m-1} x[k] y_n[k-m] + \sum_{k=m}^{N-1} x[k] y_n[k-m+N] = \lambda_n y_n[m]  
\end{align}
where $m = 0,1 .... N-1$.
The matrix is real symmetric and therefore the eigenvalues are real valued. The eigenvectors are fully real and we assume them to be $N$ periodic as given in Eq.\ref{Eq: yn_periodic}.
\begin{align}
\label{Eq: yn_periodic}
    y_n[k-m+N] = y_n[k-m]
\end{align}
This simplifies Eq.\ref{Eq:ac_summation} into 
\begin{equation}
\sum_{k=0}^{N-1} x[k] y_n[k-m] = \lambda_n y_n[m]  ; m = 0,1 .... N-1.
\end{equation}
% \textit{Almost there, the Fourier thing needs to be derived taking care of signs and conjugations !}
It may be noted that the LHS of the above equation is a case of circular correlation between $x$ and $y_n$. Therefore we can write 
\begin{equation}
y_n \circledast x = \lambda_n y_n
\end{equation}
where $\circledast$ denotes correlation.
Applying Fourier transforms, we get 
\begin{equation}
\mathcal{F}(y_n \circledast x) = \mathcal{F}(\lambda_n y_n). 
\end{equation}
Fourier transform of correlation can be written as a product, giving us
\begin{equation}
\overline{\mathcal{F}(y_n)}\mathcal{F}(x) = \lambda_n \mathcal{F}( y_n).
\end{equation}
We know that $y_n$ and $\lambda_n$ are real. Since $y_n$ is periodic, a suitable choice is sinusoidal functions.
\subsection{Cosine case}
We first assume a cosine form for $y_n$ such that $y_n[m] = cos\big(\omega_n m - \phi_i\big)$, where $\omega_n = \frac{2\pi}{N}n$. The Fourier transform of such a cosine function is a pair of delta functions as given in Eq.\ref{Eq:cos_ft}.
\begin{equation}
\label{Eq:cos_ft}
    \mathcal{F}\{cos\big(\omega_n m - \phi_i \big)\} = e^{-j\phi_i} \delta(\omega - \omega_n) + e^{j\phi_i}\delta(\omega + \omega_n)
\end{equation}
The real valued normalisation factors have been ignored for the Fourier transform as they cancel out in the subsequent steps. The correlation equation can now be written as

\begin{align}
\big[e^{j\phi_i} \delta(\omega - \omega_n) + e^{-j\phi_i}\delta(\omega + \omega_n)\big]\mathcal{F}(x) &= \lambda_n [e^{-j\phi_i} \delta(\omega - \omega_n) \notag \\ 
&\quad\quad+ e^{j\phi_i}\delta(\omega + \omega_n)].
\end{align}

Note that the above equation becomes non-zero only when $\omega = \omega_n$ or $\omega = -\omega_n$. Let us take $\omega = \omega_n$ case.
We have
\begin{equation}
e^{j\phi_i}\mathcal{F}(x)[\omega_n] = \lambda_n e^{-j\phi_i}
\end{equation}
where the $\omega_n$ component of $\mathcal{F}(x)$ is \begin{equation}
    \mathcal{F}(x)[\omega_n] = |\mathcal{F}(x)[\omega_n]| e^{j\angle\mathcal{F}(x)[\omega_n]}.
\end{equation} 
Then
\begin{equation}
e^{j\phi_i}|\mathcal{F}(x)[\omega_n]| e^{j\angle\mathcal{F}(x)[\omega_n]} = \lambda_n e^{-j\phi_i}
\end{equation}
Leading to
\begin{equation}
|\mathcal{F}(x)[\omega_n]| e^{j\big(\angle\mathcal{F}(x)[\omega_n]+2\phi_i\big)} = \lambda_n.
\end{equation}
Since $\lambda_n$ is real, we have $\angle\mathcal{F}(x)[\omega_n]+2\phi_i = 0$ or $\phi_i = -\frac{\angle\mathcal{F}(x)[\omega_n]}{2}$.
As $x$ is real, the Fourier spectrum is Hermitian and therefore it can be shown that the above relation holds for $\omega = -\omega_n$. Thus, the eigenvalue and the corresponding cosine eigenvector for a given $n$ are
\begin{align}
\lambda_n &= |\mathcal{F}(x)[\omega_n]| \quad \rm{and} \\ \nonumber
y_n[m] &= cos\Big(\omega_n m + \frac{\angle\mathcal{F}(x)[\omega_n]}{2}\Big).
\end{align}

\subsection{Sine case}
Another plausible eigenvector is $y_n[m] = sin(\omega_n m - \phi_i)$, giving

\begin{equation}
    \mathcal{F}\{sin\big(\omega_n m - \phi_i \big)\} = -j\big(e^{-j\phi_i} \delta(\omega - \omega_n) - e^{j\phi_i}\delta(\omega + \omega_n)\big).
\end{equation}
Calculating the Fourier transform and simplifying, we obtain \begin{align}
\big[e^{j\phi_i} \delta(\omega - \omega_n) - e^{-j\phi_i}\delta(\omega + \omega_n)\big]\mathcal{F}(x) &= - \lambda_n \big[e^{-j\phi_i} \delta(\omega - \omega_n) \notag \\
&\quad\quad -e^{j\phi_i}\delta(\omega + \omega_n)\big].
\end{align}

Once again, let us consider $\omega = \omega_n$ case.
We have
\begin{equation}
(e^{j\phi_i})\mathcal{F}(x)[\omega_n] = -\lambda_n (e^{-j\phi_i})
\end{equation}
Leading to
\begin{equation}
|\mathcal{F}(x)[\omega_n]| e^{j\big(\angle\mathcal{F}(x)[\omega_n]+2\phi_i\big)} = -\lambda_n.
% f e^{j(\phi_c+2\phi_i)} = -\lambda_n
\end{equation}
Once again, since $\lambda_n$ is real we have $\angle\mathcal{F}(x)[\omega_n]+2\phi_i = 0$. 
Thus, the eigenvalue and the corresponding sine eigenvector for a given $n$ are
\begin{align}
\lambda_n &= -|\mathcal{F}(x)[\omega_n]| \quad \rm{and} \\ \nonumber
y_n[m] &= sin\Big(\omega_n m + \frac{\angle\mathcal{F}(x)[\omega_n]}{2}\Big).
\end{align}

Summarising, the spectrum of an anti-circulant matrix consists of the \textit{absolute} value of the DFT spectrum of the underlying periodic pattern, with the values occurring in positive-negative pairs. The eigenvectors are sines and cosines with phases determined by the phase of the DFT. 

%%%%%%%%%%%%%%%%%%%%%%%%%%%%%%%%%%%%%%%%%%%%%%%%%%

% Don't change these lines
\bsp	% typesetting comment
\label{lastpage}
\end{document}